\documentclass[]{aa}
\usepackage{lmodern} 
\usepackage{txfonts}
\usepackage{mathtools}
\usepackage{natbib}
\usepackage{silence}
\usepackage{hyperref}
\hypersetup{
    colorlinks=true,
    linkcolor=blue,
    filecolor=magenta,      
    urlcolor=cyan,
    citecolor=blue
}

\newcommand{\feh}{\text{[Fe/H]}}

\newcommand{\enstatite}{\text{MgSiO}_{3}}
\newcommand{\forsterite}{\text{Mg}_2\text{SiO}_4}
\newcommand{\water}{\text{H}_2\text{O}}
\newcommand{\fayalite}{\text{Fe}_2\text{SiO}_4}
\newcommand{\magnetite}{\text{Fe}_3\text{O}_4}
\newcommand{\St}{{\rm St}}
\newcommand{\AU}{{\rm AU}}
\newcommand{\pluseq}{\mathrel{{+}{=}}}
\newcommand{\minuseq}{\mathrel{{-}{=}}}
\newcommand{\gas}{{\rm (g)}}
\newcommand{\solid}{{\rm (s)}}
\newcommand{\vf}{v_{\rm f}}

\newcommand{\msun}{\ensuremath{\rm M_\odot}}
\newcommand{\teff}{\ensuremath{T_{\rm eff}}}
\newcommand{\logg}{\ensuremath{\log{g}}}

\begin{document}

\title{Planet formation throughout the Milky Way:}
\subtitle{Planet populations in the context of Galactic chemical evolution}
\author{Jesper Nielsen\inst{1}
\and Matthew Raymond Gent \inst{2,3}
\and Maria Bergemann \inst{2,4}
\and Philipp Eitner \inst{2,3}
\and Anders Johansen \inst{1,5}
}

\institute{
Center for Star and Planet Formation, Globe Institute, University of Copenhagen, Øster
Voldgade 5-7, 1350 Copenhagen, Denmark
\and
Max-Planck Institute for Astronomy, 69117 Heidelberg, Germany
\and
Ruprecht-Karls-Universität, Grabengasse 1, 69117 Heidelberg, Germany
\and
Niels Bohr International Academy, Niels Bohr Institute, University of Copenhagen, Blegdamsvej 17, DK-2100 Copenhagen, Denmark
\and 
Lund Observatory, Department of Physics, Lund University, Box 43, 22100 Lund, Sweden
}
\date{Received date / Accepted date}
\abstract{As stellar compositions evolve over time in the Milky Way, so will the resulting planet populations. In order to place planet formation in the context of Galactic chemical evolution, we make use of a large ($N = 5\,325$) stellar sample representing the thin and thick discs, defined chemically, and the halo, and we simulate planet formation by pebble accretion around these stars. We build a chemical model of their protoplanetary discs, taking into account the relevant chemical transitions between vapour and refractory minerals, in order to track the resulting compositions of formed planets. 
We find that the masses of our synthetic planets increase on average with increasing stellar metallicity [Fe/H] and that giant planets and super-Earths are most common around thin-disc ($\alpha$-poor) stars since these stars have an overall higher budget of solid particles. Giant planets are found to be very rare ($\lesssim$1\%) around thick-disc ($\alpha$-rich) stars and nearly non-existent around halo stars. This indicates that the planet population is more diverse for more metal-rich stars in the thin disc. Water-rich planets are less common around low-metallicity stars since their low metallicity prohibits efficient growth beyond the water ice line. If we allow water to oxidise iron in the protoplanetary disc, this results in decreasing core mass fractions with increasing [Fe/H]. Excluding iron oxidation from our condensation model instead results in higher core mass fractions, in better agreement with the core-mass fraction of Earth, that increase with increasing [Fe/H]. Our work demonstrates how the Galactic chemical evolution and stellar parameters, such as stellar mass and chemical composition, can shape the resulting planet population.}

\maketitle
\section{Introduction}\label{sec:intro}

The composition of the central star has a huge influence on the architecture of its surrounding planetary system. It is well-established that the occurrence rate of giant planets increases strongly with stellar metallicity [Fe/H] \citep{santos2001_giantocc,johnson2010_giantocc}.\footnote{We use here the bracket notation for elemental abundance which is the logarithmic number abundance scaled with the solar abundance.} Recent work by \citet{lu2021_smallocc} has found that close-in small planets ($<$2 $M_\oplus$) are also more common around metal-rich stars than metal-poor stars, hinting towards a metallicity dependence for the occurrence of small planets as well. Smaller planets ($R$$<$$2.5\ R_\oplus$) have been observed to be more common around M-type stars than FGK-stars. This shows that other stellar properties such as mass and effective temperature also play important roles in shaping planetary systems \citep{mulders2015_SEvMstar}. 

Planets are thought to form either by either planetesimal accretion, pebble accretion, or a combination of both in protoplanetary discs around young host stars \citep{alibert2018_hybridmodel,drazkowska2022_pfreview}. It is a reasonable assumption that the chemical composition of the proto-star is the same as that of its protoplanetary disc, therefore, one may expect that the present-day host star composition, if not significantly altered by the effects of stellar physics such as atomic diffusion \citep[e.g.][]{Deliyannis1990, Proffitt1991, Deal2018}, would, to a first order, dictate the compositions of planets. There is indeed observational evidence that the iron content of planets is correlated with the photospheric metallicity [Fe/H] of their host stars \citep{adibekyan2021_comphost}. 

While a lot of effort goes into simulations of planet formation, most studies either do not include a chemical model for the compositions of planets or only include a very simplistic one, usually ignoring chemical reactions between the vapour and mineral phases. Most studies investigating the chemical compositions of planets using chemical equilibrium calculations focus on planetesimal accretion as the main formation pathway for planets \citep[e.g.][]{bond2010_pltchem,carter-bond2012_pltchem,moriarty2014_pltchem}. It has nevertheless been shown that pebble accretion could have played an important role both for the formation of gas giants \citep{bitsch2015} and for terrestrial planets \citep{johansen2021_pebSS,onyett2023_pebSS}. This means that studies combining detailed chemical calculations and pebble accretion are necessary. \citet{bitsch2020_model} used a fixed condensation temperature to simulate the formation of planets using pebble accretion around host stars with a wide variety of compositions. They found that the water contents of planets forming outside the water ice line were drastically different depending on the host star abundance, with planets being composed of $\sim$50\% water ice around metal-poor stars, [Fe/H] $=-0.4$ dex, compared to $\sim$6\% around metal-rich stars with [Fe/H] $=+0.4$ dex. Further, stars with varying Mg/Si ratios can form planets with different silicate compositions, with forsterite (Mg$_2$SiO$_4$) dominating the silicate composition for high Mg/Si ratios and enstatite (MgSiO$_3$) dominating for Mg/Si ratios near unity \citep{jorge2022_comps}. Host stars with different Fe/Mg ratios are in turn expected to form planets with different core mass fractions from 15\% around stars with low [Fe/Mg] to 40\% around stars with high [Fe/Mg] \citep{spaargaren2022_corecomp}. Clearly, it is necessary to take into account individual abundance ratios of stars to comprehensively model the compositions of the resulting planet population.

The effect of Galactic chemical evolution on the formation of planets is not yet fully understood. The Milky Way is thought to consist of the disc(s), the bulge, and the halo \citep{blandhawthorn_gerhard2016_galaxyreview}, whereby the disc is represented by the thin and thick disc stellar populations \citep{gilmore_reid1983_discs}. The halo stars display a wide range of metal ratios, $-7 \lesssim \feh \lesssim -0.5$, and may host some of the oldest stars in the Galaxy \citep{beers2005,frebel2015,Nordlander2019}. The bulge populations are reminiscent of the thick disc  \citep{melendez2008, hill2011, barbuy2018} in that the stars are $\alpha$-enhanced, but show a broader range of metallicity up to significantly super-solar values, $-1.5 \lesssim$ [Fe/H] $\lesssim +0.5$. Thick disc stars are on average richer in $\alpha$-elements compared to the thin disc stars \citep[e.g.][]{fuhrmann1998,bensby2004_discs,casagrande2011_alphafe,adibekyan2012_alpha, bergemann2014, Kordopatis2015, Hayden2015}.

The delayed production of Fe, which together with O is the main contributor to the bulk metallicity of stars, is expected to result in giant planet formation confined to the disc and the bulge, whereas halo stars may lack massive planets. Giant planets form through a core accretion process where the solid core of $\sim$10$\ M_\odot$ forms first after which it accretes a massive envelope \citep{pollack1996_planacc,lambrechts2012_giantgrowth,drazkowska2022_pfreview}. An increased metal content in the disc therefore aids the formation of giant planet cores. Indeed, \citet{bashi2022_occ_galaxy} found that thick-disc stars have a lower occurrence rate of close-in super-Earths and giant planets compared to metal-rich thin-disc stars, suggesting that the masses of planets could be different between the different stellar populations. In turn, both planet masses and planet radii have been observed to increase with host star [Fe/H], where most high-mass planets were observed around stars with higher [Fe/H] and lower [$\alpha$/Fe] \citep{narang2018_plmassfeh,swastik2022_hostchem}. Further, the compositions of the planetary building blocks have been found to vary between the thin and thick disc with the abundance of e.g. enstatite and water varying significantly between the two stellar populations as a consequence of the different abundances of $\alpha$-elements \citep{cabral2023_pbbdiscs}. Planet hosts have also been found to be enhanced in $\alpha$-elements compared to stars with similar [Fe/H] hosting no planets \citep{adibekyan2012_alpha}. This demonstrates that $\alpha$-elements can play an important role in the formation of planets. 

The occurrence rates of super-Earths and sub-Neptune sized planets have also been found to be lower for high-velocity stars than for low-velocity stars in the galaxy \citep{dai2021_velocityocc} which could be a result of differences between stellar populations as thick disc stars typically have higher velocity dispersion. \citet{santos2017_comp} used a simple stoichiometric model to calculate the iron-to-silicate mass fraction of observed planets in different stellar populations in the galaxy and found that the iron-to-silicate mass fractions and water mass fractions of planets vary significantly between the thin disc and thick disc.

In summary, recent results show that the planet populations in the main stellar populations of the Milky Way are expected to differ in both mass and chemical composition. The ongoing and upcoming exoplanet detection missions, such as TESS \citep{ricker2015_tess} and PLATO \citep{rauer2014_plato,rauer2016}, are aiming to detect and characterise thousands of more exoplanets. Specifically, PLATO \citep{Montalto2021} is expected to target sub- and super-solar metallicity stars within a significant range of masses, from $\sim0.8$ to $\sim2.2$ \msun, and consequently may cover stellar populations of all ages\footnote{We note that ages are not provided in full sky PLATO input catalogue, which is public. The range of metallicities quoted by \citet{Montalto2021} is [M/H] from $-0.9$ to $+0.3$ (their Table 3), but these are not spectroscopic estimates.}, thus probing the main parameter space of the Galactic thin and thick discs. Detailed chemical abundances are not yet available for most of the PLATO target candidates. But this will soon change with large surveys like 4MOST 4MIDABLE-HR \citep{Bensby2019} that will provide high-resolution spectroscopic for many exoplanet hosts detected by these missions. Thus understanding to what degree Galactic chemical evolution and stellar parameters, such as stellar mass, play in the formation of planets is of great importance.

In this work, we therefore aim to understand variations in planet mass and planet compositions for three different stellar populations: the halo, $\alpha$-rich stars, and $\alpha$-poor stars. In section \ref{sec:stellar_data} we present the stellar sample we use and how we separate our stellar sample into the three populations as well as our elemental abundance determination. In sections \ref{sec:disc_model} and \ref{sec:chemistry} we introduce our planet formation model and chemistry model for our protoplanetary disc respectively. In sections \ref{sec:pl_mass} and \ref{sec:results_starpop} we show our results which we discuss and summarise in sections \ref{sec:discussion} and \ref{sec:conclusion}.
\section{Stellar data}\label{sec:stellar_data}
\subsection{Stellar parameters}\label{ssec:popsep}
The abundances used in this work were derived from the analysis of the optical medium-resolution spectra collected with the Very Large Telescope (VLT) of ESO. The spectra were obtained within the Gaia-ESO large spectroscopic programme \citep{Gilmore2022, Randich2022} that aimed at a comprehensive exploration of the evolution of the disc and halo with targets broadly covering the entire range of metallicities and ages relevant to understanding the structural properties of these Galactic components. Here we employ the publicly released DR4 spectra of stars observed with the Giraffe spectrograph (resolving power R $=20\,000$). The astrophysical analysis of the spectra was carried out using the SAPP pipeline \citep{Gent2022a} designed to provide accurate estimates of atmospheric parameters and detailed chemical abundances of stars within the Bayesian framework, with the core elements based upon \citet{Serenelli2013} and \citet{Schoenrich2014}. For the detailed description of the physical inputs, algorithm, and validation on benchmark stars we refer the reader to \citet{Gent2022a} and here we briefly summarise the basic methods employed for the derivation of [Fe/H], [Mg/Fe], and Galactic population assignments of stars that are relevant in the context of this paper. 

All abundances provided by the SAPP, including iron and magnesium employed in this work, rely on the non-local thermodynamic equilibrium (NLTE) synthetic grids described in \citet{Kovalev2019}. Stellar parameters, $\teff$ and $\logg$, are primarily constrained by a combination of diagnostic techniques, including Gaia DR3 astrometry, G-band, Bp, and Rp photometry, and broad-band spectral synthesis. Here we limit the analysis to the highest-quality measurements obtained from the spectra with the signal-to-noise ratio of $>20$, accurate photometry and reliable photogeometric distances from \citet{Bailer-Jones2021} with a typical fractional uncertainty of 5{\%}, that yields $5\,325$ targets. Masses of stars were derived using the BeSPP code \citep{Serenelli2013}, which has been extensively tested on different diagnostics, including the analysis of asteroseismic targets \citep[e.g.][]{Serenelli2017}. The masses of stars in our sample range from 0.7 $\msun$ to 1.5 \msun. We note that lower-mass stars are also very valuable systems for exoplanet studies \citep[e.g][]{mulders2015_occvteff}, and indeed some of the most interesting small exoplanets are found orbiting M dwarfs \citep[e.g.][]{Luque2022}. However, our stellar sample was selected from the Gaia-ESO large spectroscopic program \citep{Gilmore2012, Randich2013, Gilmore2022, Randich2022} that had a special selection strategy optimised for a deep targeted survey of all morphological Galactic populations, including the inner and outer discs and bulge, the halo, as well clusters and streams. Because of the complex survey selection function \citep{Stonkute2016}, only very few stars in the Gaia-ESO probe the relevant mass regime and the quality of the observed data for these stars is not sufficient for the detailed analysis presented in this paper.

The Galactic population assignments follow the method of \citet{Gent2022b}, where the distinction between the thin and the thick discs is purely chemical so that low $\alpha$-stars and $\alpha$-rich stars are assumed to represent the thin and thick disc component, respectively \citep[e.g.][]{Adibekyan2011,Recio-Blanco2014, Kordopatis2015, Hayden2017}. The halo contribution was assessed through the analysis of stellar kinematics following the criteria for the accreted halo in \citet{Belokurov2018}. We note that the exact distinction between the stellar populations is not trivial, because some studies suggest the disc stars are rather mixed in the phase-space \citep[][]{Bensby2014}, whereas in other works \citep[][]{Ruchti2011} it is common to separate thin and thick discs by a kinematic cut. However, this is not critical within the scope of this paper. Here we primarily focus on the differences in planetary structure resulting from the diversity of Galactic abundance ratios and the population membership only serves as a guideline for the characteristic chemical patterns established for the disc and halo stellar populations.
\subsection{Stellar abundances}\label{ssec:abun}
\begin{figure}
    \centering
    \includegraphics[width=\linewidth]{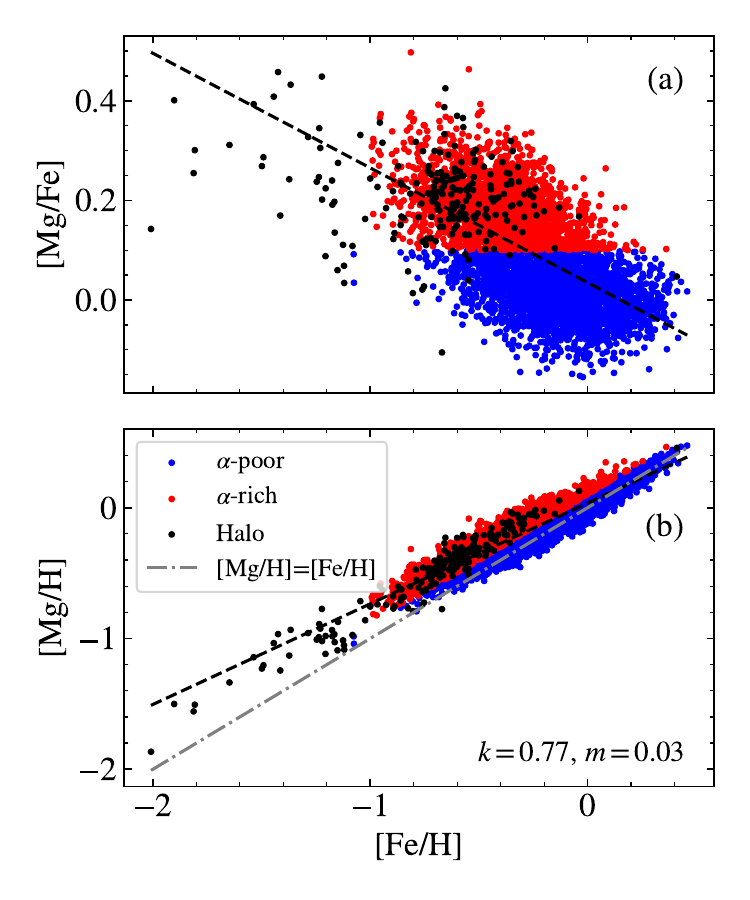}
    \caption{Stellar abundances for our stellar sample where we have colour-coded the stars belonging to each population. \textbf{(a):} [Mg/Fe] as a function of [Fe/H]. \textbf{(b):} [Mg/H] as a function of [Fe/H] for the same stars as in the top panel. We also show the linear fit between [Mg/H] and [Fe/H] for each population which was used to determine all the other elements except for He. The fit slope ($k$) and intercept ($m$) for the full sample are shown in the bottom right. In both panels, the dashed black line shows the fit when considering the entire stellar sample.}
    \label{fig:mgfe_mgh_feh}
\end{figure}

We assume that the $\alpha$-elements considered (O, Si, and S), as well as the volatile-building elements carbon and nitrogen, share the same scaling relation with [Fe/H] as magnesium and set [X/Fe] = [Mg/Fe]. For the solar abundances, we use the updated values of \citet{magg2022_solarcomp}, who recalculated the solar abundances using updated observational data as well as new and improved stellar modelling. We show the [Mg/Fe] and [Fe/H] ratios of our stars in the top panel of figure \ref{fig:mgfe_mgh_feh}, with the stellar populations marked in different colours, showing clearly how [Mg/Fe] is decreasing with increasing [Fe/H]. As seen in the bottom panel, [Mg/H] is close to linearly dependent on [Fe/H]. To be able to create data sets of synthesised stars, we extract a simple linear fit of the form $[{\rm Mg/H}] = k\feh + m$ that we then use to calculate [X/H] as a function of [Fe/H], where X are all elements other than He. The linear fit can be seen in the lower panel of figure \ref{fig:mgfe_mgh_feh}. 

To determine the abundance of He, we use the relation from \citet{leath2022_helium} who found that the He abundance for galactic globular clusters could be fitted to the form
\begin{equation}\label{eq:helium}
    Y = -0.0564\feh+0.24,
\end{equation}
where Y is the helium mass fraction of the star. This equation was fitted to globular clusters with [Fe/H] up to -0.5 so we extrapolate the fit to the values of [Fe/H] in our stellar sample.

\begin{figure}
    \centering
    \includegraphics[width=\linewidth]{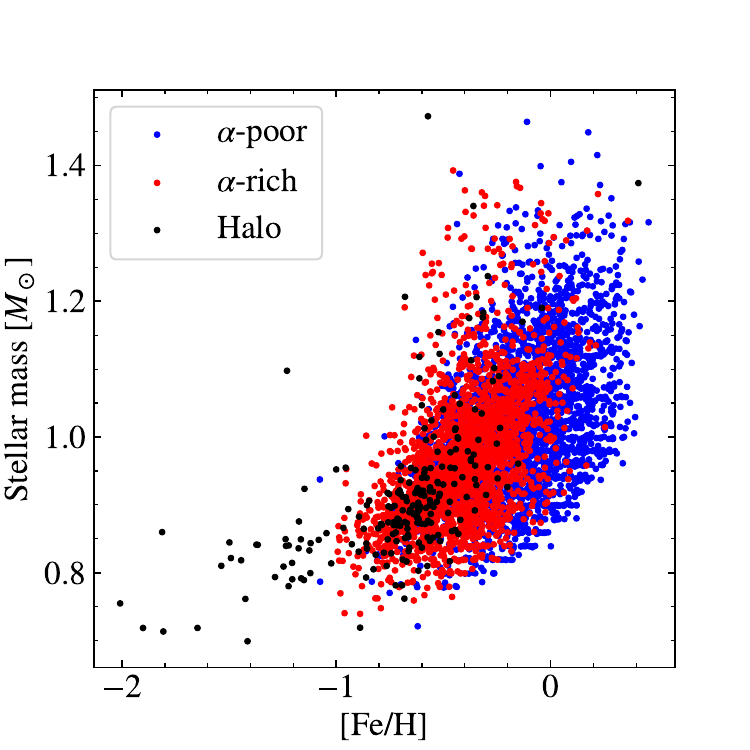}
    \caption{Stellar mass as a function of [Fe/H] for the three populations. As the lifetime of stars decreases significantly with stellar mass, there are few remaining Fe-poor stars with solar mass or above, since these stars formed early on in the evolution of the galaxy.}
    \label{fig:feh_mstar}
\end{figure}

We also show the stellar masses as a function of [Fe/H] for all three populations in figure \ref{fig:feh_mstar}. Due to the decreased lifetime of stars with higher mass, very few Fe-poor stars in the halo and $\alpha$-rich sample have super-solar masses. In contrast, young, Fe-rich stars in the $\alpha$-poor sample have an overall larger spread in stellar mass as the massive stars are not old enough to have reached the final stages of their evolution.
\section{Planet growth and disc model}\label{sec:disc_model}
\subsection{Accretion of pebbles}\label{ssec:pebble_accretion}
The classical planet growth scenario involves mutual collisions of km-sized planetesimals \citep{pollack1996_planacc}. This requires high planetesimal surface densities and is very slow beyond a few astronomical units in the protoplanetary disc due to the low accretion rate of planetesimals by protoplanets \citep{johansen_bitsch2019_planetesimalacc,lorek_johansen2022_planetesimalacc}. Considering instead the accretion of mm-cm sized pebbles gives a much shorter growth timescale of planet cores \citep{bitsch2019_growth,lambrechts2012_giantgrowth}. 

To calculate the growth rate of the protoplanet, we consider pebble accretion in the Hill regime \citep{johansen_2017_pebaccreview,schneider_bitsch2021_chemistry}
\begin{equation}\label{eq:pebacc}
    \Dot{M}_{\rm peb} = 
    \begin{cases}
        \pi R_{\rm acc}^2\frac{\Sigma_{\rm p}}{\sqrt{2\pi}H_{\rm p}}R_{\rm acc}\Omega \quad & {\rm 3D}\\
        2R_{\rm acc}\Sigma_{\rm p} R_{\rm acc}\Omega \quad & {\rm 2D},
    \end{cases}
\end{equation}
where $R_{\rm acc}$ is radius of the accretion sphere, $\Sigma_{\rm p}$ is the pebble surface density, $H_{\rm p}$ the pebble scale height, and $\Omega$ is the Keplerian frequency. We set the initial protoplanet size in this work to 0.01 $M_\oplus$. As shown by \citet{lyra2023_growth}, planetesimals can grow from the IMF emerging from the streaming instability to protoplanets with these masses for a wide range of orbital distances. For St$<$0.1, the accretion radius can be defined as
\begin{equation}\label{eq:Racc}
    R_{\rm acc} = \left(\frac{\St}{0.1}\right)^{1/3}R_{\rm H},
\end{equation}
where $R_{\rm H} = r[M/(3M_*)]^{1/3}$ is the Hill radius, $r$ is the radial distance to the star, and $M_*$ is the stellar mass \citep{lambrechts_johansen2014_pebbleflux}. The pebble scale height, $H_{\rm p}$, is defined as $H_{\rm p} = H\sqrt{\delta_{\rm t}/(\delta_{\rm t}+\St)}$ \citep{johansen2014_pp6Hp}, with $H$ denoting the scale height of the gas, $\delta_{\rm t}$ the dust diffusion coefficient. St denotes the Stokes number of the pebbles and is related to their size and density through
\begin{equation}\label{eq:St}
    \St = \frac{\pi}{2}\frac{a\rho_\bullet}{\Sigma_{\rm g}},
\end{equation}
where $a$ is the pebble size, $\rho_\bullet$ is the material density of the pebbles and $\Sigma_{\rm g}$ is the gas surface density of the protoplanetary disc \citep{drazkowska2022_pfreview}. 

In the 3D regime, the accretion radius is smaller than the pebble scale height, which means that it can only accrete parts of the pebble flux, while in the 2D regime, the protoplanet is large enough to accrete from the full pebble flux. A protoplanet accretes in the 2D Hill regime if it fulfils the criterion \citep{morbidelli2015_2d3dswitch}
\begin{equation}\label{eq:2D_crit}
    R_{\rm acc} > \frac{\pi H_{\rm p}}{2\sqrt{2\pi}}.
\end{equation}
\subsection{Temperature structure of the disc}
\label{ssec:temperature}
\begin{figure}
    \centering
    \includegraphics[width=\linewidth]{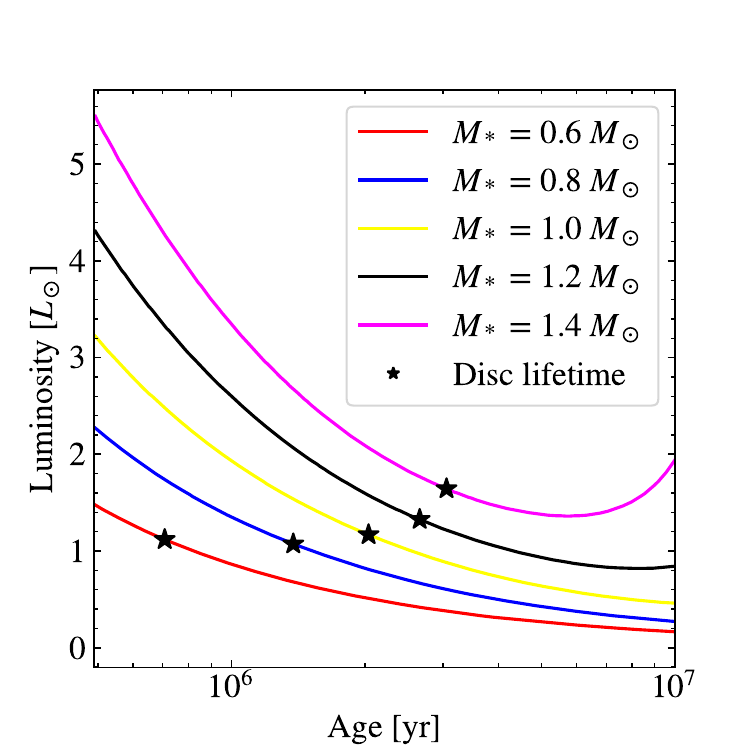}
    \caption{Luminosity isochrones over time for selected stellar masses as calculated by \citet{baraffe2015_lum}. The maximum disc lifetime for the stellar masses we consider is $\sim$3 Myr (see figure \ref{fig:disc_lifetime_sizes}). Thus the stellar luminosity, and therefore the disc temperature, decreases with time in all our models, causing ice lines and chemistry lines to migrate inwards.}
    \label{fig:lumtime}
\end{figure}

The gas scale height can be written as 
\begin{equation}\label{eq:scale_height}
    H = \frac{c_{\rm s}}{\Omega},
\end{equation}
where $c_{\rm s}$ is the sound speed of the gas, which we calculate from the ideal gas law
\begin{equation}
    c_{\rm s}=\sqrt{\frac{k_{\rm b}T}{\mu_{\rm gas} m_{\rm H}}},
\end{equation}
where $k_{\rm b}$ is Boltzmann's constant, $T$ is the temperature, $\mu_{\rm gas}$ is the mean molecular weight of the gas equal to approximately 2.3 in a protoplanetary disc with solar composition\footnote{The mean molecular weight $\mu_{\rm gas}$ varies very little with temperature and composition of the disc, so we consider it to be constant at 2.3 for all our disc models.}, and $m_{\rm H}$ is the hydrogen mass. Our temperature profile is that of a simple irradiated disc model \citep{ida2016}
\begin{equation}\label{eq:t_irr}
    T_{\rm irr} = 150\left(\frac{L}{L_\odot}\right)^{2/7}\left(\frac{M_*}{M_\odot}\right)^{-1/7}\left(\frac{r}{\AU}\right)^{-3/7},
\end{equation}
where $L$ is the luminosity of the star. We do not include any viscous heating of the disc in our model as it has been shown through nonideal magnetohydrodynamical simulations that viscous heating is inefficient compared to irradiation heating when the angular momentum is carried away by disc winds \citep{mori2021_noviscousheating}. To calculate the luminosities of the host star we make use of evolutionary tracks made by \citet{baraffe2015_lum} who calculated the luminosity of young, pre-main sequence stars, as a function of time and stellar mass. Figure \ref{fig:lumtime} shows the luminosity as a function of time for a few selected stellar masses. As the luminosity isochrones are only valid for stellar ages older than 0.5 Myr, we set the luminosity for times earlier than 0.5 Myr to be equal to the luminosity at $t=0.5$ Myr.
\subsection{Migration of the protoplanet}
\label{ssec:migration}
The migration of the protoplanet is defined through type-I migration \citep{tanaka2002_migration},
\begin{equation}\label{eq:type1}
    \Dot{r}_{\rm I} = -k_{\rm mig}\frac{M}{M_*}\frac{\Sigma_{\rm g}r^2}{M_*}\left(\frac{H}{r}\right)^{-2}v_{\rm K},
\end{equation}
where $v_K$ is the Keplerian speed at the position of the protoplanet, and $k_{\rm mig}$ is a constant prefactor. We use the fit of $k_{\rm mig}$ from 3D simulations made by \citet{dangelo_lubow2010}
\begin{equation}\label{eq:kmig}
    k_{\rm mig}=2(1.36+0.62\beta+0.43\zeta),
\end{equation}
with $\beta$ denoting the negative power-law index of the surface density, equal to 15/14 and $\zeta$ denoting the negative power-law index of the temperature, equal to 3/7. Once the protoplanet grows large enough, it transitions to type-II migration due to the formation of a gap in the disc \citep{kanagawa2018}. To include this, we follow \citet{johansen2019} and modify our migration rate to be
\begin{equation}
    \Dot{r} = \frac{\Dot{r}_{\rm I}}{1+(M/2.3M_{\rm iso})^2},
\end{equation}
where $M_{\rm iso}$ is the pebble isolation mass, defined in equation \eqref{eq:Miso}.
\subsection{Gas flux and disc lifetime}
\label{ssec:gas_flux}
The mass fluxes of pebbles and gas through the protoplanetary disc drive the mass evolution of the protoplanet. The radial accretion speed of the gas is determined by the turbulent viscosity $\nu = \alpha c_{\rm s} H$ where $\alpha$ is the viscous $\alpha$-parameter, setting the rate of angular momentum transport and is set to 0.01 throughout this work, resulting in disc lifetimes of a few Myr, expected from observations of protoplanetary discs \citep{haisch2001_disclifetime,fedele2010_disclifetime,ribas2015_disclifetime}. The gas accretion speed $u_r$ in the inner regions of the protoplanetary disc is thus
\begin{equation}\label{eq:ur}
    u_{\rm r} = -\frac{3}{2}\alpha c_{\rm s} \frac{H}{r},
\end{equation}
while the accretion speed for the pebbles $v_{\rm r}$ is 
\begin{equation}\label{eq:vr}
    v_{\rm r} = -\frac{2\Delta v}{\St+\St^{-1}}+\frac{u_{\rm r}}{1+\St^2}.
\end{equation}
Due to radial pressure support, the gas orbits at a slightly lower speed than the Keplerian speed. We can define the sub-Keplerian speed $\Delta v$ as
\begin{equation}\label{eq:deltav}
    \Delta v = -\frac{1}{2}\frac{H}{r}\chi c_{\rm s},
\end{equation}
where $\chi$ is the negative logarithmic pressure gradient and can be defined from the temperature and surface density gradients $\chi=-\partial \ln P/\partial \ln r=\beta+\zeta/2+3/2$. 

Finally, the mass fluxes of the gas can be written as
\begin{equation}\label{eq:fluxes}
    \begin{split}
        \Dot{M}_{\rm g} = -2\pi r u_{\rm r} \Sigma_{\rm g}.
    \end{split}
\end{equation}
We model the gas flux onto the star $\Dot{M}_{\rm g}$ following the $\alpha$-disc expression of \citet{hartmann1998},
\begin{equation}\label{eq:gasacc}
    \Dot{M}_{\rm g} = \Dot{M}_0\left(\frac{t}{t_{\rm s}}+1\right)^{-(5/2-\gamma)/(2-\gamma)}.
\end{equation}
Here, $\gamma$ is the power law index of the turbulent viscosity $\nu$ and can be found through $\gamma=3/2-\zeta$, where $\zeta=3/7$ is the negative logarithmic derivative of temperature. The characteristic timescale $t_s$ of the $\alpha$-disc is equal to
\begin{equation}\label{eq:ts}
    t_{\rm s}=\frac{1}{3(2-\gamma)^2}\frac{R_1^2}{\nu_1},
\end{equation}
where $R_1$ is the characteristic initial size of the disc and $\nu_1$ is the turbulent viscosity at $R_1$. We set the initial gas flux to be equal to 10 times the mass loss due to photoevaporation $\Dot{M}_{\rm w}$ from X-rays as found by \citet{owen2012_photoevap},
\begin{equation}\label{eq:photoevap}
    \begin{split}
    &\Dot{M}_{\rm w} = 6.25 \times 10^{-9}\left(\frac{M_*}{M_\odot}\right)^{-0.068}\times\\
    &\left(\frac{L_{\rm X}}{10^{30}\ \text{erg s}^{-1}}\right)^{1.14} {\rm M_\odot\ yr}^{-1},
    \end{split}
\end{equation}
where $L_{\rm X}$ is the X-ray luminosity from the star which we take from \citet{bae2013_xray},
\begin{equation}\label{eq:L_x}
    \log(L_{\rm X} [\text{erg s}^{-1}]) = 30.37+1.44\log\left(\frac{M_*}{M_\odot}\right).
\end{equation}
To get the total disc lifetime, we calculate the total gas mass lost from the disc due to both accretion onto the star and photoevaporative mass loss and set the disc lifetime $t_{\rm disc}$ to be equal to the time at which the entire gas disc has been lost. This scheme results in the gas flux going from $\sim$$10^{-7} M_\odot$ yr$^{-1}$ to $\sim$$5\times 10^{-9} M_\odot$ yr$^{-1}$ for a solar mass star which are typical values observed for solar-like stars \citep{hartmann2016}.
\subsection{Pebble flux}
\label{ssec:pebble_flux}
To calculate the pebble flux, we use a scheme similar to that of \citet{drazkowska2021_pebbleflux}. We assume that all of the dust is initially micron-sized and grows by collisions on the timescale of
\begin{equation}\label{eq:t_growth}
    t_{\rm growth} = \frac{\Sigma_{\rm g}}{\Sigma_{\rm d}} \left(\frac{\alpha_{\rm t}}{10^{-4}}\right)^{-1/3}\left(\frac{r}{\AU}\right)^{1/3}\Omega_{\rm K}^{-1},
\end{equation}
where $\Sigma_{\rm d}$ is the dust surface density, assumed to be equal to $Z_{\rm disc}\Sigma_{\rm g}$ with $Z_{\rm disc}$ the dust-to-gas ratio. We note that $\alpha_{\rm t}=10^{-4}$, is not necessarily equal to the viscous $\alpha$-parameter in equation \eqref{eq:ur} if parts of the angular momentum are carried outwards by disc winds rather than turbulent stresses. The dust is assumed to grow exponentially over time $t$ as 
\begin{equation}\label{eq:st_exp}
    \St_{\rm exp} = \St_0 \exp\left(\frac{t}{t_{\rm growth}}\right),
\end{equation}
where $\St_0$ is the initial dust size. We then consider three different processes that limit the resulting size of the dust: turbulence-driven fragmentation, fragmentation driven by differential radial drift, and the radial drift of pebbles. The maximum Stokes number from turbulence-driven fragmentation is 
\begin{equation}\label{eq:Stf}
    \St_{\rm f} = \frac{\vf^2}{3\alpha_{\rm t}c_{\rm s}^2},
\end{equation}
\citep{birnstiel2012_growth}. The Stokes number resulting from differential radial drift is in turn 
\begin{equation}\label{eq:St_df}
    \St_{\rm df} = \frac{\vf}{2\Delta v}.
\end{equation}
Finally, similar to \citet{drazkowska2021_pebbleflux}, we assume that the growth of dust needs to be 30 times faster than radial drift, resulting in a drift-limited Stokes number
\begin{equation}\label{eq:St_drift}
    \St_{\rm drift} = \frac{1}{f_{\rm drift}\eta}\frac{\Sigma_{\rm d}}{\Sigma_{\rm g}},
\end{equation}
where $f_{\rm drift} = 30$ is the drift-limiting parameter and $\eta=\Delta v/v_{\rm K}$. The resulting Stokes number is then set as the minimum between these four Stokes numbers and the initial dust size $\St_0$. We can then calculate the pebble flux from 
\begin{equation}\label{eq:pebble_flux}
    \Dot{M}_{\rm p} = 2\pi r v_{\rm r}\Sigma_{\rm d}.
\end{equation}
However, as discussed by \citet{drazkowska2021_pebbleflux}, equation \eqref{eq:pebble_flux} is only valid if the growth of dust particles is faster than their drift, allowing for a continuous supply of material. If the drift is faster, the pebble flux is limited and thus we limit the drift velocity as 
\begin{equation}
    v_{\rm r} = {\rm min}\left(\left\lvert-\frac{2\Delta v}{\St+\St^{-1}}+\frac{u_r}{1+\St^2}\right\rvert,\frac{r}{f_{\rm drift}t_{\rm growth}}\right).
\end{equation}
Similar to \citet{drazkowska2021_pebbleflux}, we keep track of the mass budget at each time step of our integration. We start by dividing the protoplanetary disc in a radial grid, calculating the initial dust mass outside of each grid cell $M_{\rm dust,0}$. At each timestep $i$, we then use the resulting pebble flux to calculate the remaining dust mass outside of each cell through
\begin{equation}\label{eq:M_remain}
    M_{\rm dust,i} = M_{\rm dust,i-1}-\Dot{M}_{\rm p}\Delta t.
\end{equation}
We then approximate the resulting dust surface density at each time step as $\Sigma_{d,i}=Z_{\rm disc}\Sigma_{\rm g}M_{\rm dust,i}/M_{\rm dust,0}$. It is important to note that in our model, the dust-to-gas ratio $Z_{\rm disc}$ of the disc is dependent on the distance to the star as we take into account the evaporation of the different chemical species in our model. For more details, see section \ref{ssec:metallicity}. Should all species be condensed out into solid form, the dust-to-gas ratio of the disc $Z_{\rm disc}$ is equal to the metal mass fraction of the star $Z$. 

Finally, we calculate the surface density of pebbles
\begin{equation}\label{eq:sigma_p}
    \Sigma_{\rm p} = \sqrt{\frac{2\dot{M}_{\rm p}\Sigma_{\rm g}}{\sqrt{3}\varepsilon_{\rm p}\pi r v_{\rm k}}},
\end{equation}
where $\varepsilon_{\rm p}$ is a measure of the efficiency of coagulation of pebbles and is set to 0.5 assuming almost perfect sticking \citep{lambrechts_johansen2014_pebbleflux}. Similarly to \citet{johansen2019}, we nominally use a dust diffusion coefficient (the diffusion equivalent to $\alpha_{\rm t}$) of $\delta_{\rm t} = 0.0001$. We also set the fragmentation velocity to 2 m/s, resulting in Stokes numbers of $\sim$0.01-0.03 at 10 AU and a pebble scale height relative to the gas $\sim$0.05-0.1, consistent with observations of dust in protoplanetary discs \citep{mulders2012_dustsett}. We explore variations in $\vf$ and $\delta_{\rm t}$ and their effect on the final masses of planets in section \ref{sec:pl_mass}. We do not consider different sizes for pebbles made up of different chemical species as, for the large protoplanets considered as starting points in our model, the differences between a single size distribution and a multiple size distribution are minimal \citep{andama2022_multispecies}. 

\begin{figure}
    \centering
    \includegraphics[width=\linewidth]{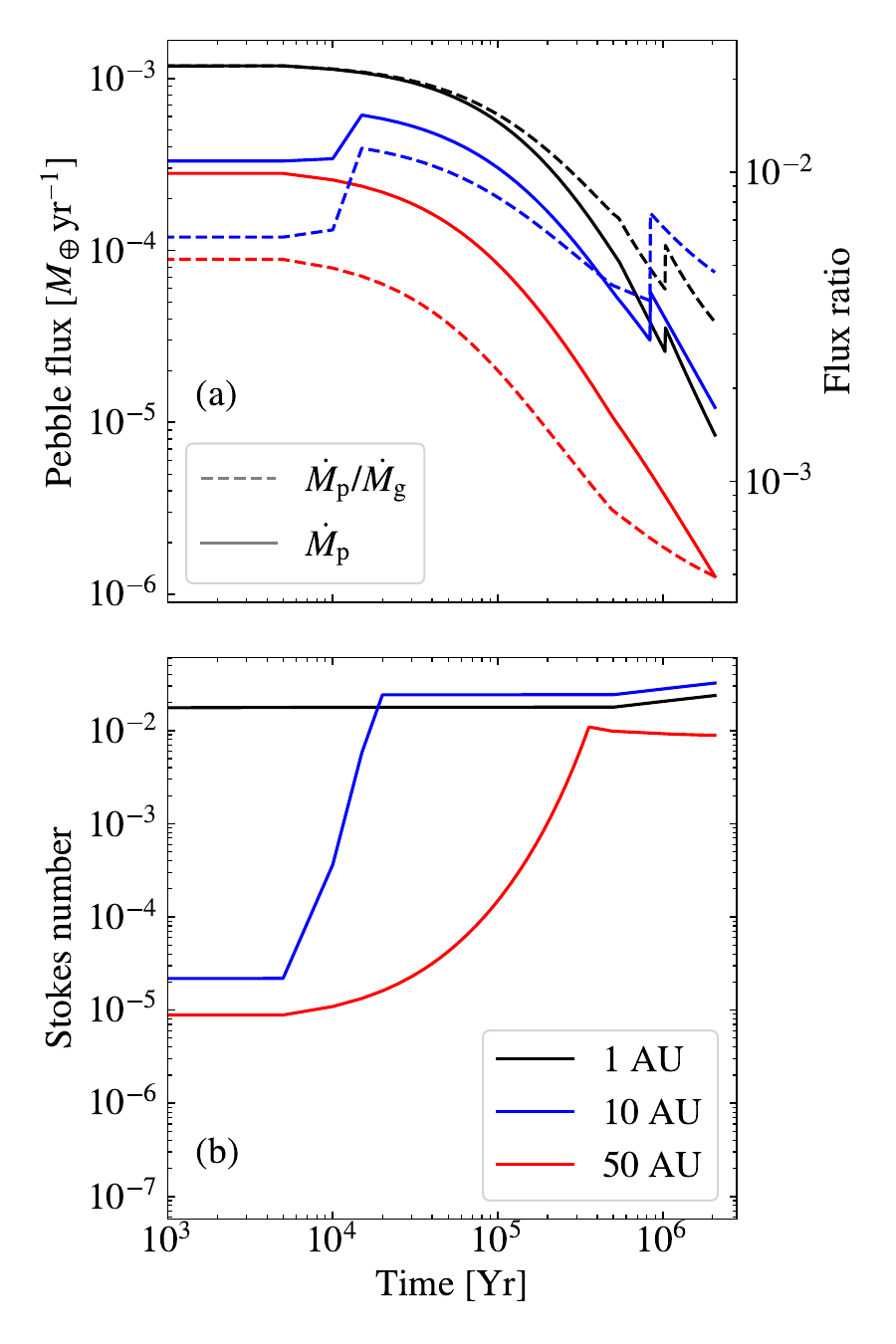}
    \caption{The evolution of the dust over time in the protoplanetary disc. The different colours represent different distances to the star. \textbf{(a):} Pebble flux and flux ratio as a function of time around a star with solar mass and composition. The fragmentation velocity here was set to 2 m/s. The flux ratio starts at a moderately high value of $\sim$0.015 but decreases rapidly after $\sim$$10^5$ years. The sharp increase in pebble flux and flux ratio at 1 Myr is caused by the CO$_2$ iceline which is traveling inwards due to the decreasing luminosity of the star. \textbf{(b):} Stokes number as a function of time for the same distances as in the top panel. The Stokes number increases fast in the inner region of the disc, growing to 0.02 before 10$^3$ years. The size of the dust is mostly regulated by fragmentation either due to turbulence or differential radial drift. Only in the outermost regions of the disc is the growth of the dust slow enough for the Stokes number to be drift-limited.}
    \label{fig:pebble_flux}
\end{figure}

We show the pebble flux and Stokes number as a function of time at different distances to a star with solar composition and mass in figure \ref{fig:pebble_flux}. Once pebbles have become large enough, the pebble flux is very high ($\sim$$10^{-3} M_\oplus/$yr) but quickly decreases as dust in the outer regions of the disc is depleted. As the star cools due to the decreasing luminosity, the CO$_2$ ice line travels inwards, eventually reaching 10 AU after $\sim$1 Myr, boosting the pebble flux slightly. The Stokes number starts initially small in the outer regions of the disc but quickly grows to $\sim$0.01-0.02. In contrast to the results of \citet{drazkowska2021_pebbleflux}, our Stokes numbers do not decrease in the later stages of the disc as our disc lifetimes are short enough for the gas and dust to be cleared before depleting enough to limit, due to long coagulation time-scales, the dust growth in the outer regions of the disc. As opposed to previous studies \citep[e.g.][]{boothilee2019_chem}, we do not track the growth of individual chemical species separately. Dust particles of similar size but different densities have different Stokes numbers and thus have different growth timescales. This difference, however, has a negligible effect on the overall growth and subsequent drift of dust in the protoplanetary disc. We also do not solve the radial transport equations for each individual species separately. We thus ignore the potential pile-up of pebbles outside of the ice lines caused by gas diffusing outwards, crossing the ice line and recondensing into pebbles \citep{boothilee2019_chem,schneider_bitsch2021_chemistry}. This effect is mostly important for planetesimal formation where the pebble pile-up could be large enough to trigger the streaming instability, forming planetesimals \citep{schoonenberg2017_planform,drazkowska2017_snowline}, something we do not consider in our model. Further, the diffusion and subsequent recondensation are most effective at relatively high values of $\alpha_{\rm t}$ in the range of $10^{-3}-10^{-2}$ \citep{schoonenberg2017_planform} which we do not consider in our model.
\subsection{Gas mass and size of the disc}
\label{ssec:gas_mass}
\begin{figure}
    \centering
    \includegraphics[width=\linewidth]{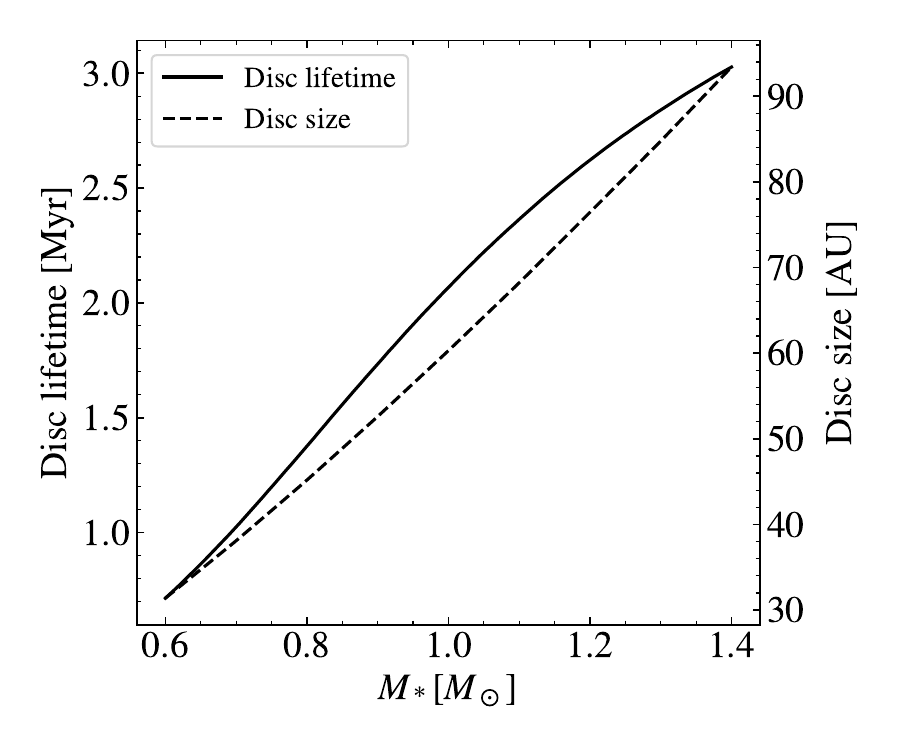}
    \caption{Protoplanetary disc lifetimes and characteristic sizes for stellar masses between 0.6 $M_\odot$ and 1.4 $M_\odot$. The disc masses were calculated from the scaling law of \citet{appelgren2020} while the lifetimes were set to the time when all gas had either been lost due to X-ray photoevaporation or accreted onto the star. The disc lifetime increases with stellar mass where the maximum lifetime is 3 Myr for $M_*=1.4\ M_\odot$, in line with observations of young stars \citep{ribas2015_disclifetime}.}
    \label{fig:disc_lifetime_sizes}
\end{figure}

We set the total gas mass of the disc to be equal to $M_{\rm gas} = 0.1(M_*/M_\odot)^{1.4}$ following the scaling relation found by \citet{appelgren2020} where we have normalised it such that a star with solar mass hosts a disc with a nominal mass of 0.1 M$_\odot$. The gas mass of the disc is related to the disc size $R_1$ by the integral
\begin{equation}\label{eq:mgas_r1}
    M_{\rm gas} = \int_{r_{\rm in}}^{R_1} 2\pi r \Sigma_{\rm g} dr,
\end{equation}
which can be solved for $R_1$ at $t=0$ to get the initial disc size as a function of the initial gas mass and accretion rate,
\begin{equation}\label{eq:R1}
    R_1=\left(r_{\rm in}^{13/14}+\frac{39k_b\alpha150L_0^{2/7}M_{\rm gas,0}}{28\dot{M}_0\sqrt{G}M_*^{9/14}\mu_{g}m_{\rm H}}\right)^{14/13},
\end{equation}
where $r_{\rm in}$ is the inner edge of the disc, set to 0.1 AU throughout this work. Figure \ref{fig:disc_lifetime_sizes} shows different lifetimes and disc sizes for stellar masses between 0.6 $M_\odot$ and 2 $M_\odot$. For solar-mass stars, we get a disc size of $\sim$60 AU and a disc lifetime of $\sim$2 Myr. Observations of young stars with ages 3-11 Myr and $M<$2 $M_\odot$ have mostly shown either no discs, transitional discs, or debris discs, while the majority of stars younger than 3 Myr host a protoplanetary disc \citep{ribas2015_disclifetime}. Therefore, lifetimes shorter than younger than 3 Myr are expected from observations for stars in our mass range. 

\subsection{Pebble isolation mass}
Once the protoplanet has grown massive enough to significantly perturb the gas around it, a barrier is formed outside of the orbit, stopping pebbles from reaching the protoplanet. The lack of heat release from infalling pebbles allows the protoplanet to cool down through radiative heat loss, which facilitates the accretion of gas \citep{lambrechts2014_gasacc}. \citet{bitsch2018} found this pebble isolation mass to be 
\begin{equation}\label{eq:Miso}
\begin{split}
    &M_{\rm iso} = 25\,M_\oplus\left(\frac{H/r}{0.05}\right)^3 \left[0.34\left(\frac{\log(\alpha_3)}{\log(\alpha_{\rm t})}\right)^4+0.66\right]\times
    \\
    &\left(1+\frac{\chi-2.5}{6}\right)\mathbf{\left(\frac{M_*}{\msun}\right)},
\end{split}
\end{equation}
through fitting 3D simulations. Here, $\alpha_3=10^{-3}$ is a constant.
\subsection{Accretion of gas onto the protoplanet}
\label{ssec:gas_accretion}
The gas accretion of a protoplanet is mainly driven by the Kelvin-Heimholtz contraction of the gas envelope due to radiative heat loss, which allows for more gas to enter the Hill sphere of the planet. This contraction occurs after the planet has reached pebble isolation mass and was found by \citet{ikoma2000} to bring in a gas at a rate of
\begin{equation}
    \left.\frac{{\rm d}M}{{\rm d}t}\right\rvert_{\rm KH} = 10^{-5}\,M_\oplus\,{\rm yr^{-1}}\left(\frac{M}{10\,M_\oplus}\right)^4\left(\frac{\kappa}{0.1{\rm m^2\, kg^{-1}}}\right),
\end{equation}
where $\kappa$ is the opacity of the envelope, which we set to 0.005 ${\rm m^2\,kg^{-1}}$. As this accretion rate scales as $\propto M^4$, it will eventually be faster than the amount of gas that can be supplied to the protoplanet. \citet{ida2018} found the gas enters the Hill sphere at a rate of
\begin{equation}
\begin{split}
    &\left.\frac{{\rm d}M}{{\rm d}t}\right\rvert_{\rm Hill} = 1.5\times 10^{-3}\,M_\oplus\,{\rm yr^{-1}}\left(\frac{H/r}{0.05}\right)^4\times
    \\
    &\left(\frac{M}{10M_\oplus}\right)^{4/3}\left(\frac{\alpha}{0.01}\right)^{-1}\left(\frac{\Dot{M}_{\rm g}}{10^{-8}M_\odot{\rm yr^{-1}}}\right)\frac{1}{1+(M/M_{\rm gap})^2},
\end{split}
\end{equation}
based on derivations by \citet{tanigawa_tanaka2016}. Should the accretion rate by contraction be slower than the Hill accretion rate, then the mass will only increase through contraction as gas cannot enter the Hill sphere faster than the contraction rate. If instead, the contraction rate is higher than the Hill accretion rate, then the gas will have enough time to contract for more gas to enter the Hill sphere and thus the Hill accretion rate sets the mass increase. Further, both of these processes are limited by the global gas accretion onto the star, since gas cannot enter the Hill radius faster than the flow of gas through the protoplanetary disc itself. Therefore, following the procedures from previous work \citep{johansen2019,ida2018} we set the gas accretion rate of the protoplanet to be the minimum of the three processes
\begin{equation}
    \left.\frac{{\rm d}M}{{\rm d}t}\right\rvert_{\rm disc} = {\rm min}\left(\left.\frac{{\rm d}M}{{\rm d}t}\right\rvert_{\rm KH},\left.\frac{{\rm d}M}{{\rm d}t}\right\rvert_{\rm Hill},\Dot{M}_{\rm g}\right).
\end{equation}
\section{Chemistry model}\label{sec:chemistry}
To calculate the number densities of chemical species in the disc, we partition the disc into temperature ranges where different chemical species dominate. This forms what we refer to as \textit{chemistry lines} at which certain reactions take place, similarly to ice lines which are temperatures at which different elements or species sublimate. We use the temperatures at which these chemical reactions occur as reported by \citet{fegley2000} who based these on calculations of equilibrium chemistry. Should the pebble drift timescale be shorter than the chemical reaction timescale, refractory species would be lost to the star before having time to chemically react with vapour in the disc. However, chemical reaction timescales for the temperatures present in our disc model are short enough for chemical equilibrium to be a valid assumption \citep{semenov2010_chemtime}.
\subsection{Carbon}
\label{ssec:carbon}
Our carbon model is similar to that of \citet{oberg2011_carbon} which in turn is based on observations of gas and ice in protoplanetary discs as well as ISM grain modelling \citep{draine2003_carbongrains,pontoppidan2006_COobs}. Throughout the entire disc, the carbon species are distributed as shown in table \ref{tab:carbon}. The number abundances for CO and CO$_2$ are set as the minimum between a fraction of the carbon or the amount of oxygen available. The remaining carbon exists in turn in refractory carbon grains. Following \citet{oberg2011_carbon}, the carbon grain condensation temperature has been set to a high temperature. We use the condensation temperature of graphite as reported by \citet{lodders2003_temps}. All of these carbon species are present throughout the entire disc in all temperature ranges.
\subsection{Nitrogen}
\label{ssec:nitrogen}
For Nitrogen, we follow  \citet{schneider_bitsch2021_chemistry} and set the number abundances of NH$_3$ and N$_2$ to be equal to 0.1 N/H and 0.9 N/H, respectively, as N$_2$ is expected to be the major nitrogen carrier \citep{cleeves2018_nitrogen,bosman2019_nitrogen}. In turn, NH$_3$ has been observed to only have a minor contribution to the nitrogen abundance in protoplanetary discs with at most 10\% of the nitrogen present as NH$_3$ ice \citep{cleeves2018_nitrogen,pontoppidan2019_nitrogen,bosman2019_nitrogen}.
\begin{table}
    \centering
    \caption{Number abundance of carbon- and nitrogen-bearing species}
    \label{tab:carbon}
    \begin{tabular}{l|l}
        Species & Number abundance\\\hline
        CO (condensation at 20 K) & min(0.65C/H, O/H) \\
        CO$_2$ (cond. at 70 K) & min(0.2C/H, (O/H-CO)/2) \\
        C (cond. at 626 K) & C-CO-CO$_2$ \\
        NH$_3$ (cond. at 90 K) & 0.1 N/H \\
        N$_2$ (cond. at 20 K) & 0.9 N/H
    \end{tabular}
\end{table}
\subsection{Refractory minerals}
\label{ssec:minerals}
Since we do not consider viscous heating, our discs will not become hotter than $\sim$1000 K. We nevertheless still include refractory chemistry up to forsterite condensation at 1354 K \citep{lodders2003_temps} and ignore minerals more refractory than this. We show the resulting abundances of each species in table \ref{tab:number_abundances}. Below, we discuss the chemical reactions we take into account going from high to low temperatures.

The condensation of forsterite (Mg$_2$SiO$_4$) at 1354 K can occur through the reaction
\begin{equation}\label{eq:forsterite}
    2 \text{Mg}\gas + \text{SiO}\gas + 3 \water\gas = \forsterite\solid+3\text{H}_{2}\gas.
\end{equation}
In equilibrium, forsterite will take up either all the water vapour, silicon, or magnesium depending on which element there is the least of. SiO$\solid$ will retain the remaining excess of oxygen or silicon. Depending on the abundance ratios, there might be some free silicon or magnesium left.

A few tens of degrees (1316 K) below the forsterite condensation \citep{fegley2000,lodders2003_temps}, enstatite (MgSiO$_3$) condenses through the reaction
\begin{equation}\label{eq:enstatite}
    \forsterite\solid+\text{SiO}\gas+\water\gas=2\enstatite\solid+\text{H}_{2}\gas.
\end{equation}

To calculate the abundance of enstatite, we use the abundance of silicon monoxide, forsterite, and water from the previous temperature range. Should all forsterite be used up, leaving some SiO$\gas$ left, it will react with some of the water, forming SiO$_{2}\solid$ instead through the reaction
\begin{equation}\label{eq:SiO2}
    {\rm SiO}\gas + \water\gas = {\rm SiO}_{2}\solid.
\end{equation}
At about 710 K, H$_2$S$\gas$ begins to corrode the metallic iron in the disc, forming FeS$\solid$ through the reaction
\begin{equation}\label{eq:FeS}
    {\rm H}_2{\rm S}\gas + {\rm Fe}\solid = {\rm FeS}\solid.
\end{equation}

\begin{figure}
    \centering
    \includegraphics[width=\linewidth]{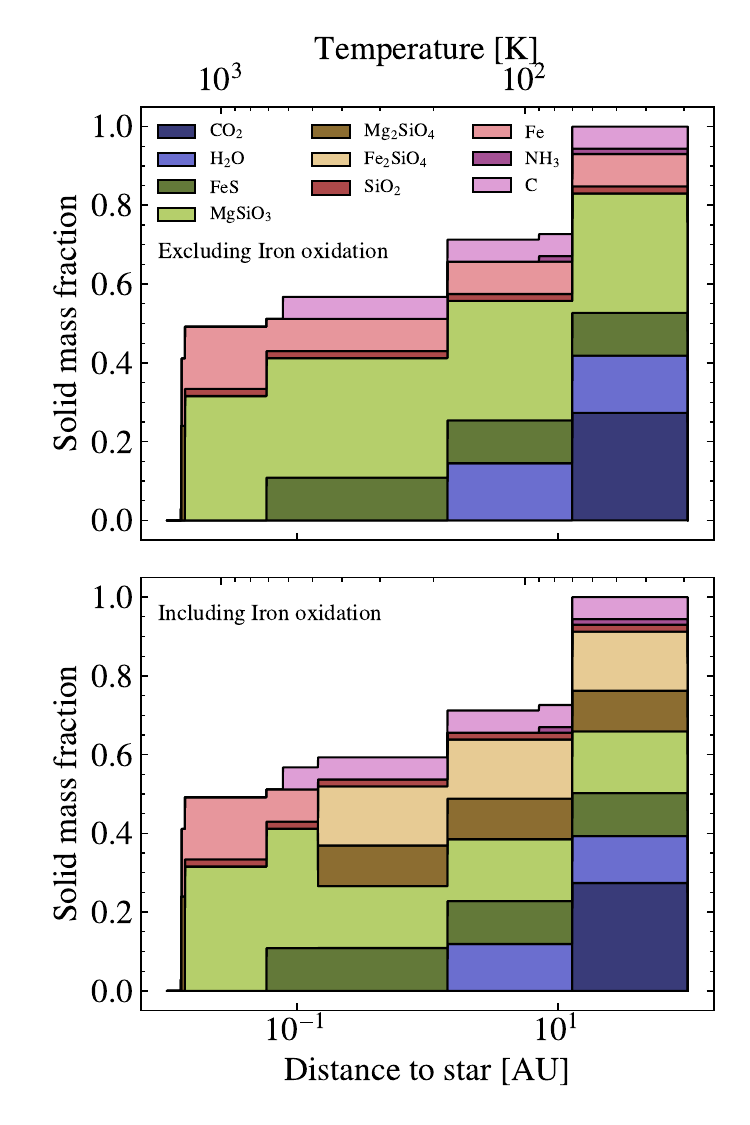}
    \caption{Mass fraction of solids in the protoplanetary disc around a star with solar mass and composition for both iron oxidation models. The mass fractions are normalised such that all considered species add up to one. FeS condensation occurs around 0.05 AU while fayalite condensation occurs at 0.15 AU. The condensation of water vapour into water ice happens just outside of 1 AU in the disc while CO$_2$ condenses out outside of 10 AU.}
    \label{fig:solid_fraction}
\end{figure}

We show the mass fraction of solids in the disc in figure \ref{fig:solid_fraction} normalised such that the sum of all elements shown adds up to unity when they are all condensed out. We also show how the number densities are calculated based on the stellar abundances for all temperature brackets in table \ref{tab:number_abundances}.

\subsection{Iron oxidation}
\label{ssec:iron_ox}
At approximately 480 K, any remaining metallic iron reacts with enstatite and water, forming fayalite (Fe$_2$SiO$_4$) through the reaction
\begin{equation}\label{eq:fayalite}
\begin{split}
    &2\text{Fe}\solid+2\enstatite\solid+2\water\gas=\\
    &\forsterite\solid+\fayalite\solid+2\text{H}_{2}\gas.
\end{split}
\end{equation}
Thus, there is a switch in the abundances of enstatite and forsterite where the more abundant enstatite is partially (or fully, depending on the amount of free Fe) lost to forsterite.

At 370 K any remaining metallic iron will be consumed into the condensation of magnetite through the reaction
\begin{equation}\label{eq:magnetite}
    3 \text{Fe}\solid+4\water\gas = \magnetite\solid + 4 \text{H}_{2}\gas.
\end{equation}
As discussed by \citet{grossman2012_ironoxidisation}, iron oxidation is unlikely to occur freely in the disc during the lifetime of the protoplanetary disc due to the difficulty of diffusing metallic iron into already condensed forsterite and enstatite. Instead, it is only possible for iron to oxidise if the nucleation of metallic iron is halted, causing supersaturation of iron, ultimately leading to fayalite production being significantly increased. Such an environment could exist e.g. inside already formed planetesimals or by taking into account suppression of homogeneous nucleation of iron until very high supersaturations \citep{johansen2022_mercury}. To test the effects that iron oxidation has on the final planet population, we test two models: one which does not allow for oxidation of iron where all iron exists as metallic iron Fe(s) or FeS(s), and one where we allow iron to oxidise to form fayalite and magnetite. All of our results presented exclude iron oxidation unless stated otherwise.

\begin{table*}
    \caption{Number densities for all species and all considered temperature ranges. We make use of $\minuseq$ and $\pluseq$ in order to show change compared to the previous, hotter temperature bracket. Any species not shown compared to the hotter temperature bracket remains unchanged. We set the water abundance available for forsterite condensation to be equal to all the oxygen not present in carbon volatiles or SiO. The initial SiO number density available for forsterite condensation is set to the minimum between silicon and the oxygen not trapped in carbon volatiles.}
    \label{tab:number_abundances}
    \centering
    \begin{tabular}{l|l}
    Species & Number density relative to hydrogen \\\hline
    \textbf{1316-1354 K} & \\
    Mg$_2$SiO$_4$ (solid) & min($\frac{1}{2}$Mg/H,SiO,$\frac{1}{3}$H$_2$O) \\
    SiO (gas) & min(Si/H,O/H - CO - 2CO$_2$) - Mg$_2$SiO$_4$ \\
    Mg (gas) & Mg - 2Mg$_2$SiO$_4$ \\
    H$_2$O (gas) & O/H - 2CO$_2$ - CO - 3Mg$_2$SiO$_4$\\\hline
    \textbf{710-1316 K} & \\
    MgSiO$_3$ (solid) & min(2Mg$_2$SiO$_4$,2SiO,2H$_2$O) \\
    SiO$_2$ (solid) & min(SiO,H$_2$O)\\
    Mg$_2$SiO$_4$ (solid) & $\minuseq \frac{1}{2}$MgSiO$_3$ \\
    SiO (gas) & $\minuseq$ (SiO$_2+\frac{1}{2}$MgSiO$_3$) \\
    H$_2$O (gas) & $\minuseq$ (SiO$_2+\frac{1}{2}$MgSiO$_3$)\\\hline
    \textbf{480-710 K} & \\
    FeS (solid) & min(Fe,H$_2$S) \\
    Fe (solid) & $\minuseq$ FeS \\
    H$_2$S (gas) & $\minuseq$ FeS \\\hline
    \textbf{370-480K} & \\
    Fe$_2$SiO$_4$ (solid) & min(Fe/2,$\frac{1}{2}$MgSiO$_3$,$\frac{1}{2}$H$_2$O) \\
    Mg$_2$SiO$_4$ (solid) & $\pluseq$ Fe$_2$SiO$_4$ \\
    MgSiO$_3$ (solid) & $\minuseq$ 2Fe$_2$SiO$_4$ \\
    Fe (solid) & $\minuseq$ 2Fe$_2$SiO$_4$ \\
    H$_2$O (gas) & $\minuseq$ 2Fe$_2$SiO$_4$ \\\hline
    \textbf{$<$370 K} & \\
    Fe$_3$O$_4$ (solid) & min($\frac{1}{3}$Fe,$\frac{1}{4}$H$_2$O) \\
    H$_2$O (condensation at 180 K) & $\minuseq$ 4Fe$_3$O$_4$ \\
    Fe (solid) & $\minuseq$ 3Fe$_3$O$_4$
    \end{tabular}
\end{table*}
\subsection{Temperature of the planet atmosphere}
\label{ssec:atmo_temp}
During pebble accretion, the gravity of the planet will attract a hydrostatic envelope in smooth contact with the protoplanetary disc at the Hill radius. Following \citet{piso_youdin2014}, we calculate the temperature at the radiative-convective boundary (RCB) for a planet-atmosphere system embedded in a protoplanetary disc, which is defined as the radius at which the temperature profile goes from isothermal (equal to the temperature in the disc at a specific radius) to adiabatic. Should the sublimation temperature of any species be lower than the temperature at the RCB during pebble accretion, we work under the assumption that those pebbles will evaporate during the infall onto the planet and be lost back into the disc. We show detailed derivations of the temperature at the RCB in appendix \ref{asec:T_rcb}.
\subsection{Calculating the dust-to-gas ratio}
\label{ssec:metallicity}
\begin{figure}
    \centering
    \includegraphics[width=\linewidth]{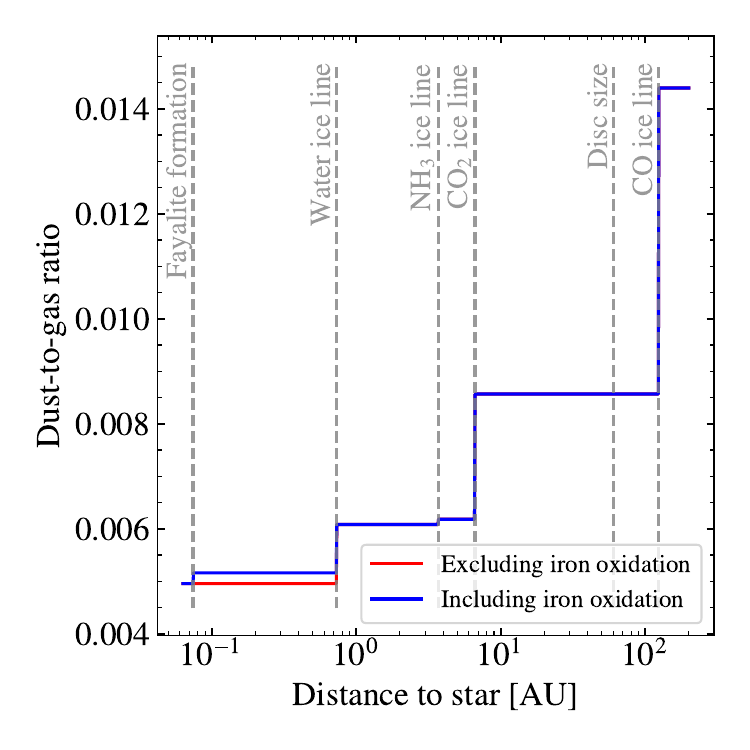}
    \caption{Dust-to-gas ratio as a function of distance to a solar-mass and solar-composition star at $t=2$ Myr. In the colder regions, further away from the star, the dust-to-gas ratio increases due to the condensation of species at their ice lines. We also show the calculated disc size for the star. The condensation temperature for CO is low enough for it to never condense out in the disc but we show it here for clarity.}
    \label{fig:metallicity}
\end{figure}

In order to determine the dust-to-gas ratio of the disc, we consider all species that have condensed into solid form for each distance $r$ from the star. The resulting ratio can then be expressed as
\begin{equation}
    Z_{\rm disc} = \sum_i \frac{n_i\mu_i}{1+n_{\rm He}\mu_{\rm He}},
\end{equation}
where $n_i$ is the number abundance of each solid species and $\mu_i$ is their atomic mass. We show in figure \ref{fig:metallicity} the dust-to-gas ratio as a function of distance from a solar-metallicity star for the model including iron oxidation and the model without. The significant boost to the dust outside of 1 AU is due to the water ice line, while the increase in the amount of solid material outside of 10 AU is attributed to the CO$_2$ ice line. Importantly, we note that the dust-to-gas ratio of the disc will, for the inner region, be less than the metallicity of the host star as many volatile species have sublimated. Comparing the nominal model without iron oxidation to the model including iron oxidation, the only difference occurs between the fayalite line and the water ice line. A small amount of water vapour is used during fayalite condensation, trapping oxygen in the refractory fayalite instead of water vapour. Outside the water ice line, this difference disappears as all water vapour freezes out.
\subsection{Calculating the amount of accreted material}
\label{ssec:M_acc}
At each time step, the number abundance of all the chemical species at the location of each protoplanet is calculated according to the above scheme. The total accreted mass of each species $i$ is then found through
\begin{equation}\label{eq:species_mass}
    \Dot{M}_i = \frac{n_i\mu_i}{\sum_j n_j\mu_j}\Dot{M}_{\rm acc},
\end{equation}
where the index $j$ is summed over all species, $n$ is the number abundance, $\mu$ is the molecular mass, and $\Dot{M}_{\rm acc}$ is the mass accreted during this timestep. During pebble accretion, only solid species are accreted and thus $j$ is indexed over all condensed species. During gas accretion, only sublimated species are considered.
\section{Planetary populations around stars of different metallicities and masses}
\label{sec:pl_mass}
We show in figure \ref{fig:MR_all} the full final planet populations around three synthesised stars with [Fe/H] = -0.5, 0, and 0.5 with metal mass fractions $Z=$ 0.006, 0.016, and 0.041, respectively. Here, we have initialised 10$^4$ protoplanets at linearly sampled times between $t=10^4$ yr and $t=0.95\,t_{\rm disc}$. The starting positions of the protoplanets are generated from a log-uniform distribution between 0.5 and the initial disc size $R_1$. The median of the initial positions is $\sim$4 AU while the 16th and 84th percentiles are $\sim$1 AU and $\sim$15.5 AU respectively. We categorise the planets depending on their composition where we define planets with $>$15\% gas as gaseous planets (red) while water-rich planets (blue) have a water mass fraction (wmf) $>$10\%. Any planet not falling under either of these categories we classify as a rocky planet (black). We also categorise planets according to their mass into giants ($>$10 $M_\oplus$), super-Earths (2-10 $M_\oplus$), and Earth-mass planets (0.05-2 $M_\oplus)$. All categories and their definitions can be found in table \ref{tab:cat_def}. We do not consider planets below 0.05 $M_\oplus$ as they have not grown sufficiently to be categorised. We integrate the models over the entire protoplanetary disc lifetimes with a time step of $10^3$ years. Different time steps between $10^2$ years and $10^4$ years were tested and we found negligible differences for time steps smaller than $10^3$ years.
\begin{table}[t]
    \centering
    \caption{Our categorisation of planets used throughout this work. We categorise planets either according to their mass (Giant, Super-Earth, or Earth-mass) or composition (Gaseous, Water-rich, or Rocky) such that the total number of planets is equal to the sum of each category in either the composition-based categories or the size-based categories. In all categories, we do not consider any bodies below 0.05 $M_\oplus$.}
    \label{tab:cat_def}
    \begin{tabular}{l|p{6cm}}
    Name & Definition \\\hline
    \textbf{Composition} & \\
    Gaseous & $M_{\rm Gas} > 0.15\ M_{\rm pl}$ \\
    Water-rich & WMF $>$10\%\\
    Rocky & Not gaseous or rocky \\\hline
    \textbf{Mass} & \\
    Giant & $M_{\rm pl} > 10\ M_\oplus$ \\
    Super-Earth & $2 < M_{\rm pl}/M_\oplus < 10$ \\
    Earth-mass & $M_{\rm pl} < 2\ M_\oplus$
    \end{tabular}
\end{table}
\subsection{Changing stellar metallicity}
\label{ssec:stellar_metallicity}
\begin{figure}
    \centering
    \includegraphics[width=\linewidth]{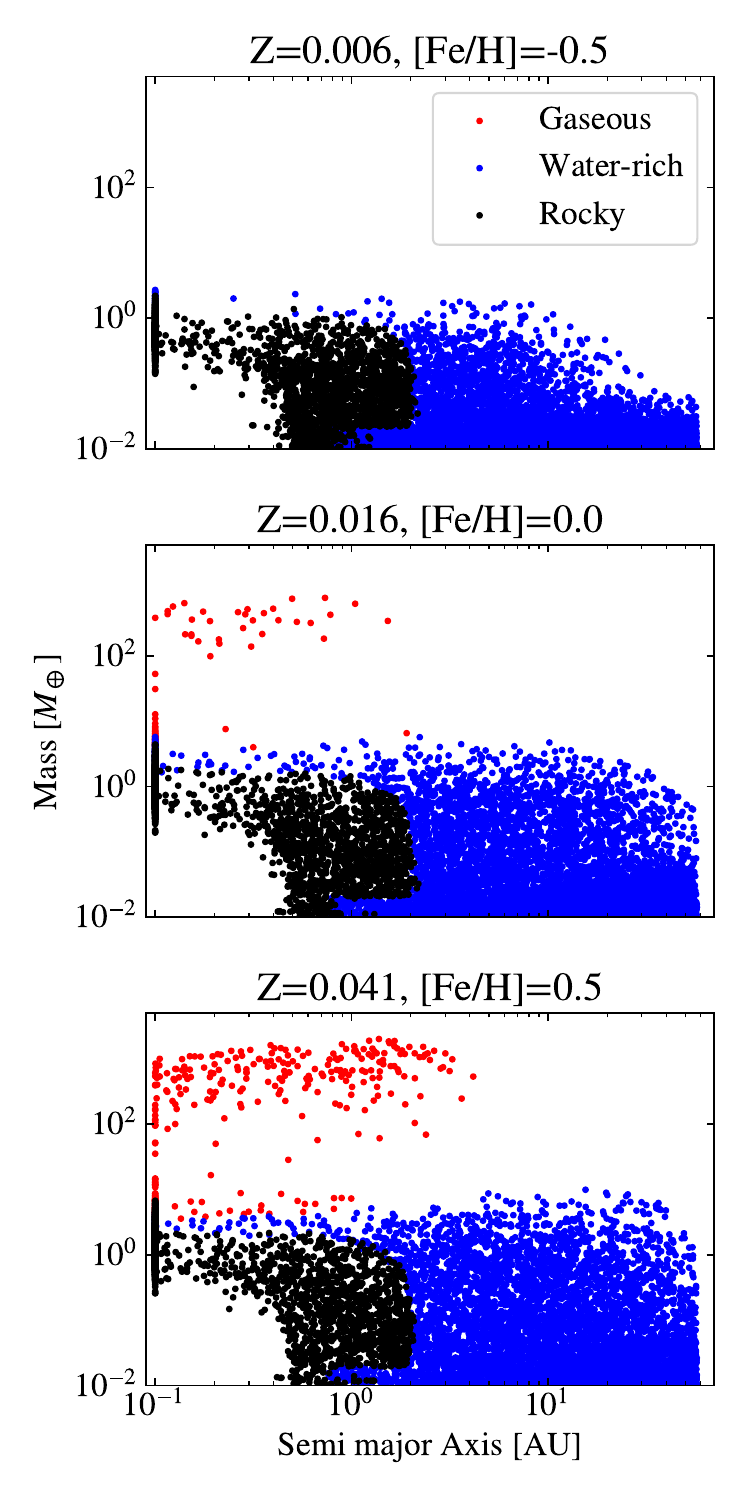}
    \caption{Resulting planet population for three solar-mass stars of varying composition. At low metallicity, it is difficult to form any planets beyond a few AU in the protoplanetary disc, so very few water-rich planets above 1 $M_\oplus$ form in that case. For solar metallicity and above (middle and bottom panels), we start to form water-rich planets up to around a few $M_\oplus$ far out in the disc. We note that due to our abundance scaling of the remaining elements, the middle panel has a slightly higher metallicity of 0.016 for solar [Fe/H] compared to the solar metallicity of 0.014.} 
    \label{fig:MR_all}
\end{figure}

As can be seen in the top panel in figure \ref{fig:MR_all}, it is very difficult to form any water-rich planets above 1 $M_\oplus$ around stars with low metallicities. This is mainly caused by a low pebble flux outside the water ice line. For solar and super-solar metallcities, Earth-mass planets that are water-rich are formed out to 50 AU. More massive water-rich planets will reach the pebble isolation mass and quickly grow to form gas giants. The effects of migration are clearly seen in the water-rich planets found close to the star. We also find that giant planets are difficult to form around stars with solar [Fe/H] and below. In order for a protoplanet to reach runaway gas accretion, it must reach a core mass of $\gtrsim$10 $M_\oplus$. As the pebble isolation mass increases further out in the disc, it is only possible for planets to grow to a large enough mass at distances of $\gtrsim$10 AU, after which the planets migrate inwards.\footnote{For detailed growth tracks of giant planets, see e.g. \citet{johansen2019}} Assuming solar metallicity, the pebble flux at those distances is high enough for the protoplanets to reach pebble isolation mass and form a small number of Jupiter-sized planets. In the high-metallicity case, it is very easy to form giant planets up to $\sim$$10^3\ M_\oplus$ with up to a few AU in final semi-major axes. We note however that since we do not include the effects of gravitational interactions between planets in our model, we cannot compare the final semi-major axes of planets to the observed semi-major axis distribution of planets.

\begin{figure}
    \centering
    \includegraphics[width=\linewidth]{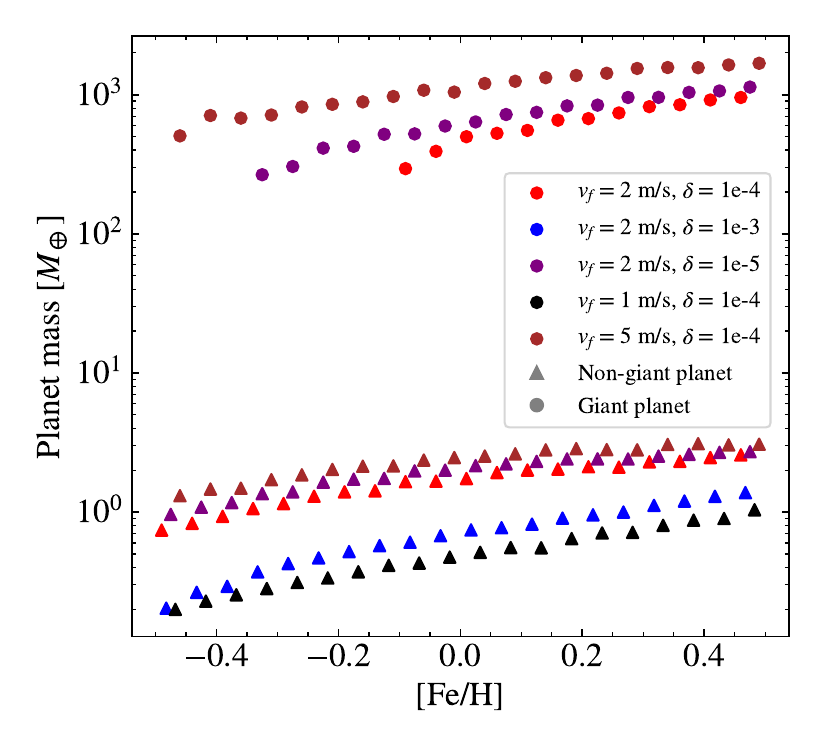}
    \caption{Mass weighted mean planet masses for giant planets (filled circles) and non-giant planets (triangles) as a function of [Fe/H] for different values of St and $\delta_{\rm t}$. We bin the stars in [Fe/H] with widths of 0.05 dex. Increasing the [Fe/H] clearly results in higher planet masses.}
    \label{fig:mpl_vs_feh}
\end{figure}

We then synthesise 100 stars with solar mass and [Fe/H] linearly spaced between -0.5 and 0.5 in order to understand the full effect on the planet populations when varying [Fe/H]. We initialise 2000 protoplanets for each star, using similar initial conditions as for the three stars in figure \ref{fig:MR_all}. In figure \ref{fig:mpl_vs_feh} we show the mass-weighted mean planet mass for giant planets and non-giant planets as a function [Fe/H] for different variations of fragmentation velocity $\vf$ and $\delta_{\rm t}$.
\\
\\
For all cases, both the giant planet mass and the rocky/water-rich non-giant planet masses increase with increasing [Fe/H]. For our nominal case of $\vf=2$ m/s, $\delta_{\rm t}=0.0001$, the mean giant planet mass increases from $\sim$$300\ M_\oplus$ to about $\sim$$900\ M_\oplus$ while the mean non-giant planet masses increases from $\sim$$0.8\ M_\oplus$ to $\sim$$3\ M_\oplus$. This trend agrees well with observational evidence that the mass of observed giant planets and radii of smaller planets increases with [Fe/H] \citep{petigura2018_occ_vs_feh,swastik2022_hostchem}. The masses for small planets can be straightforwardly calculated from a simple mass-radius relationship. Increasing radii with [Fe/H] for small planets thus also indicates increasing masses with [Fe/H] \citep{chen_kipping2017_mr}. For solar metallicities and below, it becomes very difficult for the protoplanets to grow fast enough to become gas giants. This agrees well with observations of metal-poor stars where giant planets are rare \citep{santos2001_giantocc,johnson2010_giantocc,buchhave2012_plmassvsfeh,thorngren2016_giantsfeh}. We discuss forming giant planets around stars of solar metallicity in more detail in section \ref{ssec:giants_around_suns}. 

For $\delta_{\rm t} = 10^{-3}$, the masses of rocky/water-rich non-giants are reduced by almost an order of magnitude around Fe-poor stars compared to our nominal value of $\delta_{\rm t}=10^{-4}$. This effect is smaller for stars with higher [Fe/H] due to the increased amount of solids that still allow for efficient planetary growth. Further, giant planet formation is completely inhibited for all metallicities when the turbulent diffusivity is high. A larger dust diffusion coefficient results in stirring of the pebble scale-height ($H_{\rm p}/H \approx 0.3$ for $\delta_{\rm t}=10^{-3}$, compared to $H_{\rm p}/H = 0.1$ in the nominal case) and therefore the protoplanets are stuck longer in the slow 3D Hill growth regime.

For a lower turbulent diffusion coefficient of $\delta_{\rm t} = 10^{-5}$, the pebbles are much more settled ($H_{\rm p}/H = 0.03$) resulting in a slight increase in terrestrial planet masses but more significantly, the lower turbulent diffusion allows for the formation of giant planets around stars with [Fe/H] down to -0.3. 

Increasing the fragmentation velocity to 5 m/s has a very similar effect to decreasing $\delta_{\rm t}$ for the non-giant planets as increased $\vf$ results in larger Stokes numbers. With a high enough fragmentation velocity, it is possible to form giant planets down to [Fe/H] = -0.4. When lowering the fragmentation velocity to 1 m/s, we significantly inhibit planet growth to the point that no giant planet is formed at any metallicity value considered and planets are barely able to grow at all for [Fe/H] below 0. Lower fragmentation velocities significantly inhibit the growth of dust which leads to very low Stokes numbers. This not only causes less settling of the pebbles, but the pebbles are also more difficult to accrete as the accretion radius scale as St$^{1/3}$. 
\subsection{Changing stellar mass}
\label{ssec:stellar_mass}
\begin{figure*}
    \centering
    \includegraphics[width = \linewidth]{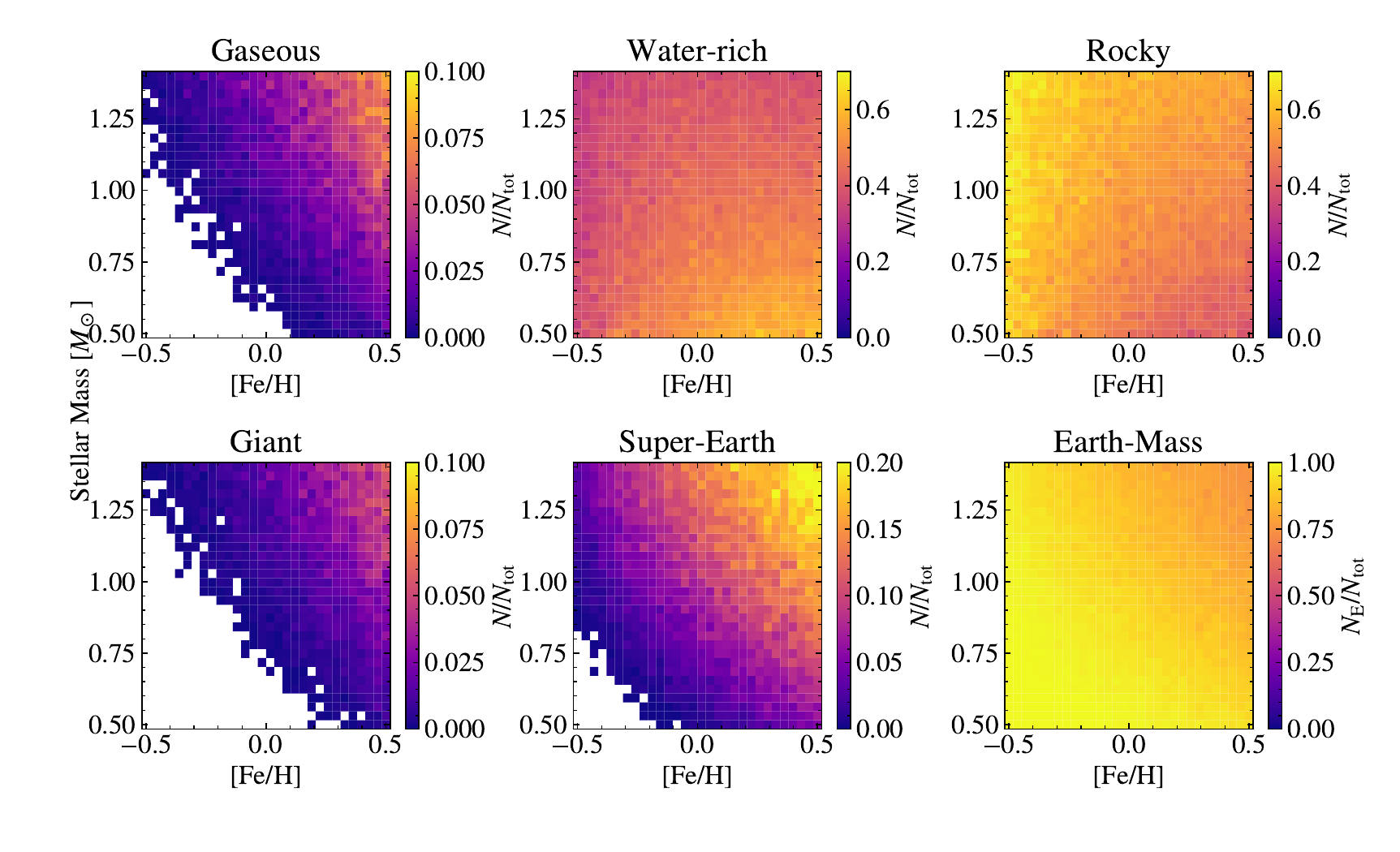}
    \caption{Fraction of different types of planets formed around stars in a grid with varying [Fe/H] and $M_*$. In order to form giant planets and super-Earths we require either a high metallicity or a high stellar mass. Further, water-rich planets are more common around stars with higher metallicities and stellar mass as these stars can sustain a high enough pebble flux outside the water ice line to form water-rich planets there. Earth-mass planets dominate for all metallicities and stellar masses as they do not require a high pebble flux in order to form.}
    \label{fig:compfrac_feh_mass}
\end{figure*}

As both the size and mass of the protoplanetary disc increase with stellar mass, a higher pebble flux will be sustained in protoplanetary discs around stars with higher mass. In order to investigate the effect of stellar mass on planet formation in detail, we synthesised a grid of 30$\times$30 stars in the [Fe/H]-$M_*$ plane with [Fe/H] ranging from -0.5 to 0.5 and $M_*$ ranging from 0.5 to 1.4 $M_\odot$. We then count the number of planets formed according to the categorisation shown in table \ref{tab:cat_def}. The resulting planet fractions for each star can be seen in figure \ref{fig:compfrac_feh_mass}. 

Giant planets are formed at [Fe/H] down to -0.5 for $M_*$$\sim$1.4 $M_\odot$ and there is a clear increase in both the fraction of giant planets and super-Earths for stars with greater mass and higher [Fe/H]. Water-rich planets are more common around low-mass, high [Fe/H] stars. High-metallicity stars have an increased pebble flux far out in the disc due to the higher dust-to-gas ratio which allows for the formation of more planets outside the water-ice line. High-mass stars form less water-rich planets than low-mass planets given the same [Fe/H] while the fraction of gaseous planets increases with increased stellar mass. Massive stars are so efficient at forming giant planets that the formation of rocky and water-rich planets is limited as they instead grow massive enough to attract a significant gaseous envelope and thus reach runaway accretion, becoming giant planets. 

\citet{bashi2022_occ_galaxy} investigated the planet occurrence for main-sequence\footnote{They do not explicitly target main-sequence stars but they only include stars in the temperature range 4500-6500K with log $g>$ 4, meaning that they do exclude any possible giants.} hosts with different stellar parameters and found that the average number of close-in super-Earths per star decreases with increasing stellar effective temperature. This has been shown previously as well by other authors \citep{howard2012_occvteff,mulders2015_SEvMstar}. Our results indicate that it is easier to form both giant planets and super-Earths around stars with higher mass (and thus higher effective temperature), which would contradict these findings. Stellar hosts with higher stellar mass (and thus higher effective temperature) are able to more easily form multiple massive planets in the disc due to a higher disc mass and thus longer lifetime and higher pebble flux. Such systems could quickly become unstable leading to the ejection of one or several planets in the system \citep{mustill2017_instability,buchhave2018_gpscatt} which would explain the observed decrease in occurrence with effective temperature. Further, gap formation caused by the formation of a giant planet could halt the inwards migration of super-Earths and thus lead to an observed anti-correlation between the occurrence of super-Earths and stellar mass \citep{vandermarel2021_discs_planets}.
\section{Planetary populations for different galactic stellar populations}
\label{sec:results_starpop}
Using our categorisation of planets in table \ref{tab:cat_def}, we can now investigate the resulting planet populations in each of our stellar populations described in section \ref{sec:stellar_data}. We initialise 2000 protoplanets around each star using the same procedure as described in section \ref{sec:pl_mass} using a fragmentation velocity of $\vf = 2$ m/s and a dust diffusion coefficient $\delta_{\rm t} = 0.0001$.
\subsection{Planet masses}
\label{ssec:pl_mass}
\begin{figure}
    \centering
    \includegraphics[width=\linewidth]{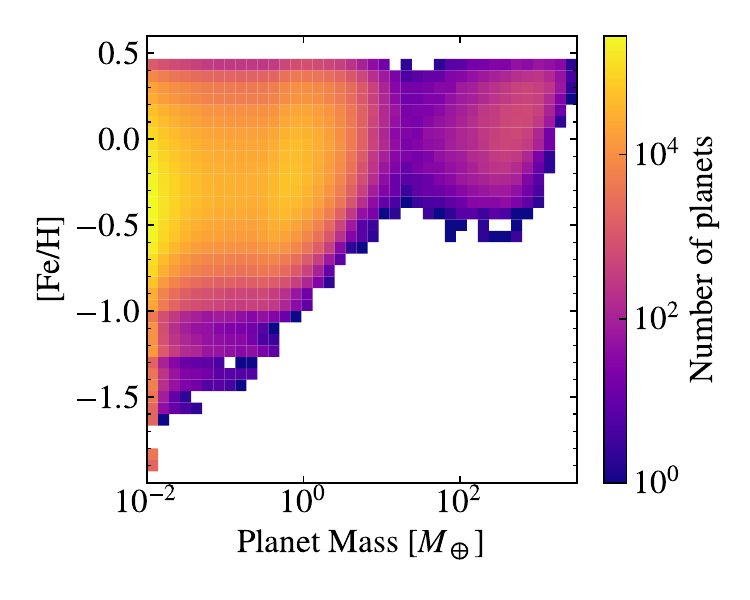}
    \caption{2D-histogram of the number of planets as a function of [Fe/H] and planet mass using our combined stellar samples of alpha-poor, alpha-rich and halo stars. Only stars with high enough [Fe/H] form massive planets, while the distribution of host star [Fe/H] is more spread out for small planets, similar to the results of \citet{buchhave2012_plmassvsfeh} and \citet{swastik2022_hostchem}. There is a clear valley in planet mass between 10 and 100 $M_\oplus$ caused by the runaway accretion of gas.}
    \label{fig:feh_plmass}
\end{figure}

We show the final planet mass for all our initialised protoplanets versus their host star [Fe/H] as a 2D histogram in figure \ref{fig:feh_plmass}. Massive planets form predominantly around stars with solar or super-solar [Fe/H], although there is some formation of massive planets around stars down to [Fe/H] = -0.5. Non-giant planets form around stars with a wide range of [Fe/H], a relation which has been shown observationally as well \citep{buchhave2012_plmassvsfeh,swastik2022_hostchem}.
\subsection{Occurrence rates}
\label{ssec:occ_rates}
\begin{figure}
    \centering
    \includegraphics[width=\linewidth]{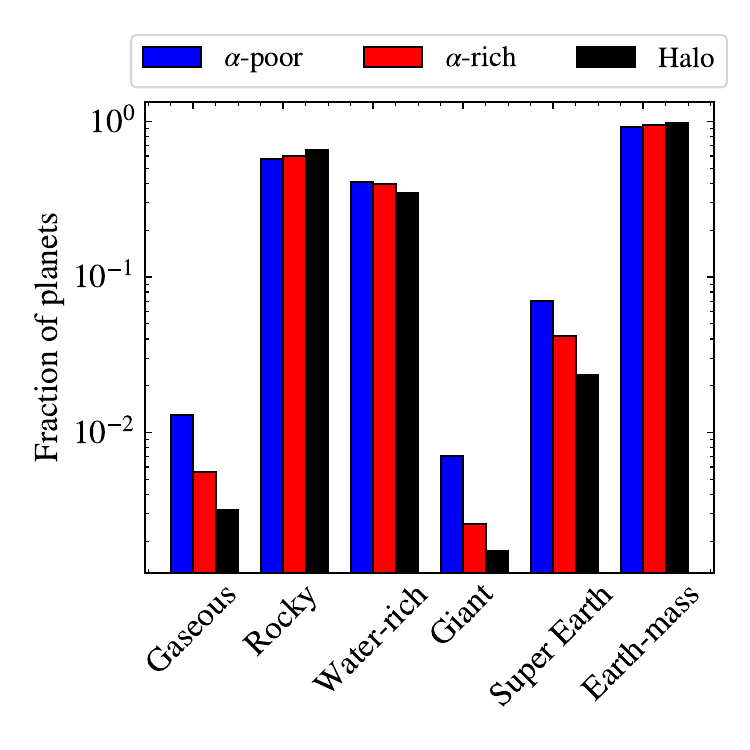}
    \caption{Fraction of planets sorted into our six planet categories for each stellar population. The total number of planets is normalised for each stellar population such that all composition-based categories (Gaseous, Rocky, Water-rich) add up to 1 and all mass-based categories (Giant, Super-Earth, Earth-mass) add up to 1. The most common planet formed is rocky in composition and Earth-like in mass. Due to the overall high [Fe/H] of $\alpha$-poor stars, water-rich planets and super-Earths are most common around these stars.}
    \label{fig:occ_bar}
\end{figure}

In figure \ref{fig:occ_bar} we show the fraction of planets belonging to each of our categories, normalised such that all mass-based categories add up to 1 and all composition-based categories add up to 1 in each stellar population. Unsurprisingly, planets that are rocky in composition and Earth-like in mass are the most common in all three populations. Due to the overall higher metallicities of $\alpha$-poor, they form slightly more water-rich planets and super-Earths than $\alpha$-rich and halo stars. The relative water abundance in protoplanetary discs is expected to decrease with increasing [Fe/H] in both iron oxidation models. When excluding iron oxidation, the mass fraction of water abundance decreases slightly due to the increased abundance of Fe. When iron oxidation is included, the water abundance in the protoplanetary decreases more significantly due to the increasing availability of iron for the condensation of fayalite (and magnetite if the iron abundance is high enough), which becomes a major oxygen carrier. However, given the higher metal content in the protoplanetary discs around stars with higher [Fe/H], more planets are able to form further out in the disc, beyond the water ice line, which counteracts the slight decrease in water content in the protoplanetary disc, forming more water-rich planets. Gaseous planets and giant planets, in turn, are slightly more common around $\alpha$-poor stars compared to $\alpha$-rich stars. 

\begin{figure*}
    \centering
    \includegraphics[width = \linewidth]{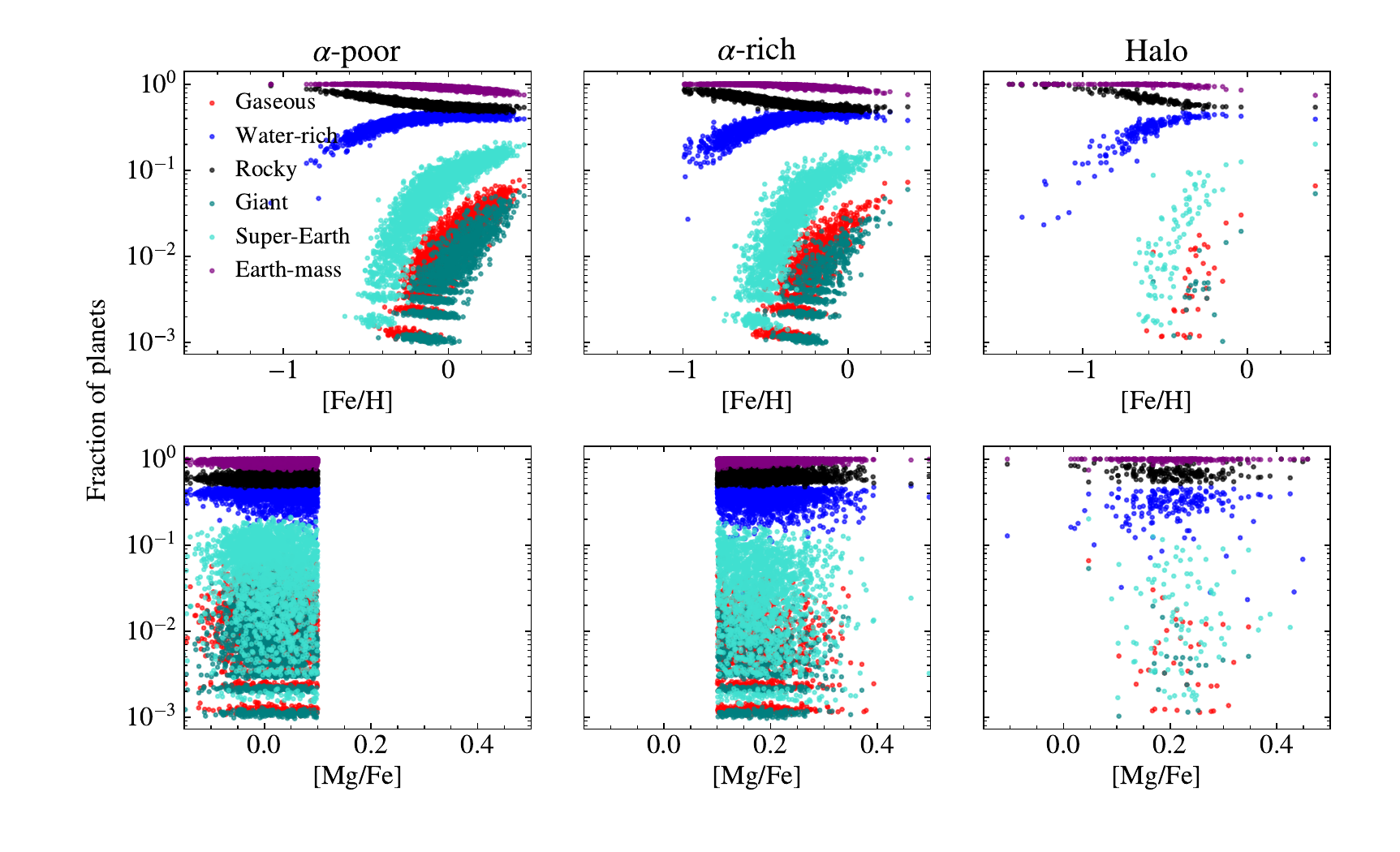}
    \caption{Occurrence rates for the different planet categories as a function of [Fe/H] (top row) and [Mg/Fe] (bottom row). The occurrence rates of giant planets increase as expected with [Fe/H], so most of the $\alpha$-rich stars and almost all of the halo stars form no giant planets due to the overall lower metallicities of those populations. Water-rich planets are common ($\sim$50\%) around both $\alpha$-poor and $\alpha$-rich stars with [Fe/H]>-0.5. There is no strong relation in the fraction of giant planets or super-Earths with [Mg/Fe], although [Mg/Fe]-poor stars are more likely to form more giant planets and super-Earths since they are typically [Fe/H]-rich.}
    \label{fig:comp_mgfe_feh}
\end{figure*}

In order to understand how the planet population is distributed with respect to both [Mg/Fe] and [Fe/H] of the host stars, we show the fraction of different types of planets in figure \ref{fig:comp_mgfe_feh}. Clearly, the fraction of water-rich planets increases for higher [Fe/H], which means that we can find water-rich planets around both $\alpha$-poor and $\alpha$-rich stars, while only a few halo stars with high metallicities form water-rich planets. The fraction of water-rich planets is not significantly changed with [Mg/Fe] within each population. The fraction of water-rich planets around $\alpha$-rich is slightly decreasing with increasing [Mg/Fe] but this is mainly caused by stars with low [Mg/Fe] being [Fe/H]-poor and thus not forming as many water-rich planets. The fraction of giant planets, gaseous planets, and super-Earths are mostly flat with [Mg/Fe] for all populations with a large scatter, caused by the large variation of [Fe/H] for stars with fixed [Mg/Fe]. For stars with similar [Fe/H], we expect the fraction of giant planets and super-Earths to increase slightly with increasing [Mg/Fe] as the increased Mg-content increases the dust-to-gas ratio in the protoplanetary disc and thus aids slightly in forming more massive planets.
\subsection{Effect of iron oxidation on the core mass fraction}
\label{ssec:iron_ox_cmf}
\begin{figure*}
    \centering
    \includegraphics[width=\linewidth]{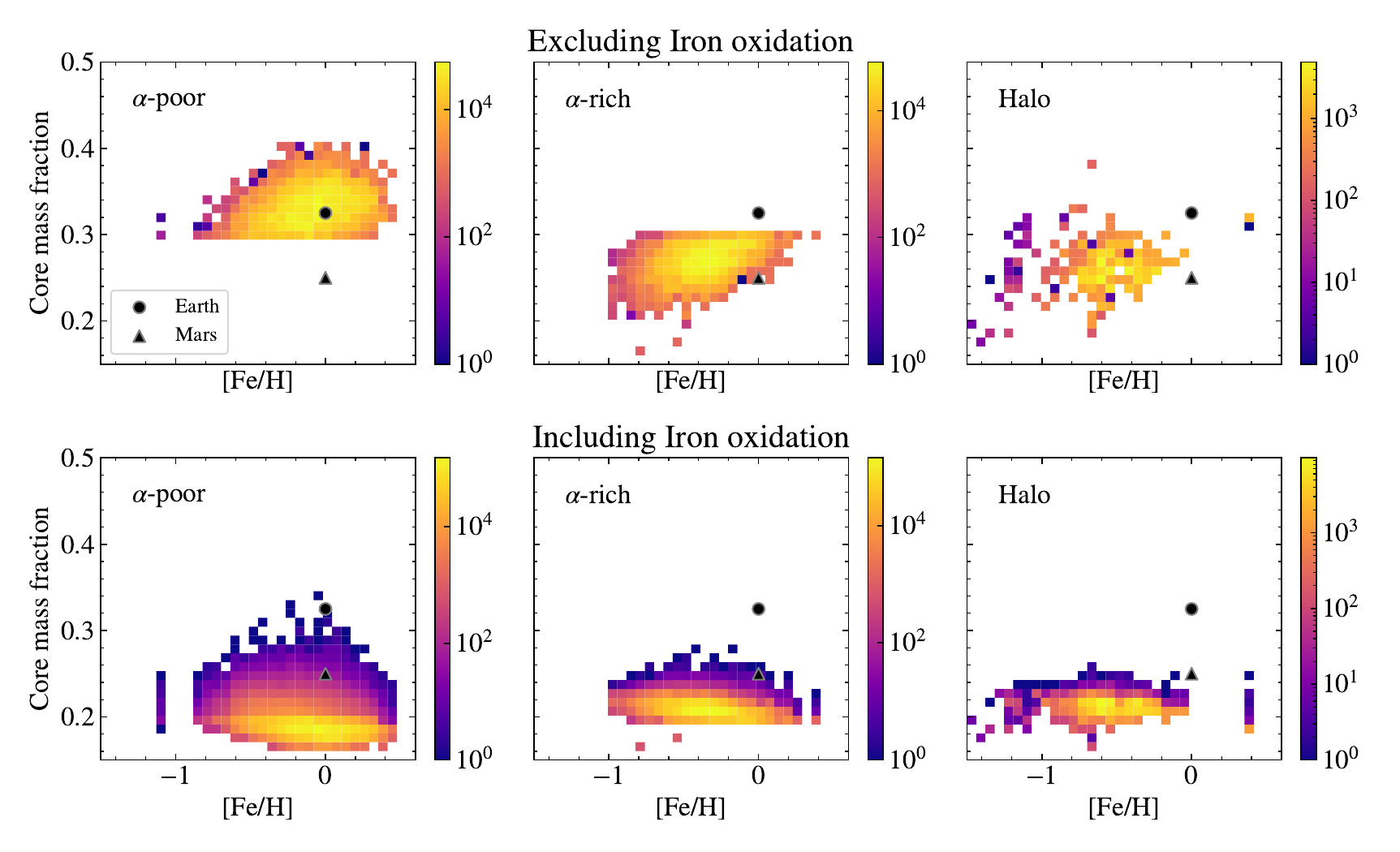}
    \caption{2D histogram of the core mass fraction as estimated by the mass of metallic iron and FeS compared to the total mass excluding water. We show results both excluding (top row) and including (bottom row) iron oxidation. When iron oxidation is excluded, the peak in the distribution in the $\alpha$-poor sample increases from $\sim$0.2 to $\sim$0.35. Further, core mass fractions increase with increasing [Fe/H] in both the $\alpha$-poor and $\alpha$-rich sample. In contrast, when including iron oxidation, the core mass fractions are relatively flat with [Fe/H] in the $\alpha$-poor and halo samples, while it decreases in the $\alpha$-rich sample as the amount of iron available to become oxidised increases. The core mass fractions of planets around halo stars remain largely unaffected by iron oxidation as [Mg/Fe] is mostly flat with [Fe/H] for halo stars.} 
    \label{fig:cmf_freeiron}
\end{figure*}

As discussed in section \ref{ssec:iron_ox}, the oxidation of iron by water in the protoplanetary disc may not be realistic. Iron oxidation affects the core mass fractions of the planets since fayalite and magnetite are lithophiles. While we do not consider a detailed interior model for our planets, we can estimate a core mass fraction by considering the mass of the species that dominate in the core (metallic iron and FeS) and compare these to the mantle, excluding any volatiles such as water. We show 2D histograms of the core mass fractions of planets and host star [Fe/H] for both iron models in figure \ref{fig:cmf_freeiron}. 

When including iron oxidation, all metallic iron is used up in the formation of fayalite and magnetite, which results in only FeS contributing to the core mass fraction. As the amount of FeS is set by the amount of S in the disc, the relative abundance of FeS compared to fayalite and magnetite decreases with [Fe/H] due to [S/Fe] decreasing with increasing [Fe/H]. Without iron oxidation, metallic iron exists freely in the disc and the amount of metallic iron increases with [Fe/H]. This causes an increase in core mass fraction with increasing [Fe/H] for planets around both $\alpha$-rich and $\alpha$-poor stars, consistent with previous work \citep{adibekyan2021_comphost}. The core mass fractions of halo stars stay mostly flat with [Fe/H] due to the flat relation in [$\alpha$/Fe] with [Fe/H]. 

We also show the core mass fractions of both Earth and Mars as calculated by \citet{wang2018_earthcmf} and \citet{khan2022_marscmf} respectively. When excluding iron oxidation, most of the planets formed have a similar core mass fraction to that of Earth, while no planets formed have a core mass fraction comparable to that of Mars. In comparison, when including iron oxidation, most core mass fractions are significantly lower than that of Earth and Mars, as all iron remaining after the condensation of FeS has already been oxidised. Mars' low core mass fraction can be explained by the oxidation of accreted metallic iron in contact with liquid water, leading to a lower core fraction and higher FeO fraction for Mars in the model of \citet{johansen2023_cmftheory}. We refer the reader to that paper for a detailed discussion.

Sulfur is assumed to be completely reacted with metallic iron, forming FeS, in agreement with previous work \citep{kama2019_sulfur}. The FeS is then partitioned into the core, resulting in a sulphur mass fraction of $\sim$6-7\% in the formed planets. However, metal-silicate partitioning experiments \citep{suer2017_sulphur} have recently found that the Earth's core cannot contain more than $\sim$2 wt\% of sulfur. Such sulfur-poor cores are not formed in this work, which shows that further modelling of the chemistry and partitioning of sulfur is needed.
\section{Discussion}\label{sec:discussion}
\subsection{Model limitations}
\label{ssec:model_discussion}
In our disc model, we have set the value of the angular momentum transport coefficient $\alpha$, which drives the gas flux onto the star, to 0.01, motivated by observations of disc lifetimes. However, other observational constraints on $\alpha$ estimates have been proposed to imply that $\alpha$ is in the range of  $3\times10^{-4}-3\times10^{-3}$ \citep{trapman2020_alpha,rosotti2023_alphaobs}. From equation \eqref{eq:R1}, it is clear that for a fixed disc mass and initial mass accretion rate, a smaller angular momentum transport coefficient would result in significantly smaller initial disc sizes. In contrast, \citet{najitabergin2018_alpha} found that such low values of $\alpha$ and small initial disc sizes could not accurately match the observed disc sizes of Class 0/1 objects. Such small disc sizes would also cause the dust to be drained from the disc incredibly fast, inhibiting planet growth. Further, \citet{appelgren2023_alpha} found good matches with the observed properties of protoplanetary discs and models with $\alpha = 0.01$. In order to achieve initial disc sizes comparable to observations with lower values of $\alpha$, either the discs would have to be more massive or the mass accretion rate have to be significantly lower. Such low accretion rates do not match observations \citep{hartmann2016} and would cause the disc lifetimes to be much longer than observed. Clearly, the choice of $\alpha$ is sensitive and can impact the final planet population significantly and further observational constraints are necessary in order to accurately model the angular momentum transport in protoplanetary discs.

We tested different lower limits of injection times for the planet embryos and found negligible differences in the overall planet population if the earliest injection time was below $10^4$ years. While the pebble fluxes are high in the disc early on, the total integrated dust mass at 1 AU at $10^4$ years is only 10 $M_\oplus$ meaning that planet growth is limited at this time. Further out in the disc, the integrated dust mass is even lower due to the limited dust growth and lower pebble flux far out in the disc. As seen in figure \ref{fig:pebble_flux}, the pebble flux decreases quickly after $10^4$ years meaning that planet embryos injected after $10^4$ years are less likely to grow to significant sizes. Indeed, setting the earliest injection time to $10^5$ yr and later significantly decreased the probability of forming giant planets. Further, we found no differences when choosing a log-uniform versus linear sampling of injection times.

Our choice of log-uniform sampling of injection location for all discs relies on the assumption that stellar parameters such as metallicity and stellar mass do not affect the formation of planet embryos. We choose a log-uniform sampling instead of linear uniform sampling so as to not limit the formation of small planets in the inner disc. Planetesimal formation is mainly driven by the streaming instability \citep{johansen2014_pp6Hp} and requires a high dust-to-gas ratio \citep[][section 3.1.3 and references therein]{drazkowska2022_pfreview}. Proposed locations for the triggering of the streaming instability are just outside of snowlines in the protoplanetary disc \citep{drazkowska2017_snowline}, the locations of which are dictated by the stellar mass. However, due to the limited mass range of our stellar sample, we do not consider this effect as the locations of the snowlines vary very little across our stellar mass range. Further, high stellar metallicities could result in planetesimal formation being possible further out in the disc due to higher dust-to-gas ratios, something we do not consider. A comprehensive planetesimal formation model would be necessary to capture these effects and is beyond the scope of this work.
\subsection{Putting planet formation in the context of galactic chemical evolution}\label{ssec:GCE}
Our results show that stars with sub-solar [Fe/H] form mostly low-mass rocky planets, while stars with super-solar metallicity form a more diverse population of planets with a mixture of water-rich planets, giant planets, and super-Earths. We also find that an increasing metallicity allows for the formation of more water-rich planets due to the increased pebble flux. The fraction of giant planets and super-Earths decreases slightly with increasing [Mg/Fe], although there is a large scatter due to a wide variation of [Fe/H] for a given [Mg/Fe]. Observed planet masses have been found to be increasing with decreasing [$\alpha$/Fe] as well as with decreasing host star ages \citep{swastik2022_hostchem,swastik2023_ages}. This further shows that as stars get more and more chemically enriched over time, the formation of more massive planets is enhanced as well. 

The sizes and masses of planetary cores are thought to have a great impact on the structure and potential habitability of planets \citep{noack2014_int_hab,dyck2021_core_hab}. Understanding to what degree iron actually oxidises in the disc is therefore very important in order to understand the structures and compositions of planets throughout the galaxy. Without iron oxidation, the core mass fractions of most planets formed are $\sim$35\% in the $\alpha$-poor sample, approximately the same as the core mass fraction of Earth \citep{seager2007_cmfearth,wang2018_earthcmf}. Including iron oxidation lowers the core mass fractions of most planets around $\alpha$-poor stars to approximately 20\%. We note that our neglect of Ni in the core should lead to an underestimate of the core mass fraction by approximately 10\%. Further, the core mass fractions of planets decrease with increasing [Fe/H] when we include iron oxidation while they are both higher as well as increasing with increasing [Fe/H] without iron oxidation. Assuming that iron does not oxidise in the disc results in core mass fractions increasing with [Fe/H] and thus over time as stars in the galaxy get more and more Fe-rich.
\subsection{Comparison with previous work}\label{ssec:compare}
\citet{bitsch2020_model} developed a similar chemical model of protoplanetary discs as this work, but they assumed that the number density of all chemical species\footnote{Ices as well as refractory minerals.} remained constant with temperature, ignoring condensation of refractory minerals as a result of chemical reactions between gas phase molecules and that existing minerals react with vapour and gas. Further, they also assumed that iron oxidises in the protoplanetary disc, forming $\magnetite$ and Fe$_2$O$_3$. They found that the water mass fraction of planets forming outside of the water ice line decreases with increasing host star [Fe/H], due to a decreasing [O/Fe] and [C/O] with increasing [Fe/H]. This results in water mass fractions in protoplanetary discs ranging from $\sim$50\% to $\sim$6\%, for host star [Fe/H] of -0.5 and 0.5, respectively. We can find similar results in our model including iron oxidation. As [C/O] is constant in our stars, we find that the water mass fraction of planets decreases less in the same range of host star [Fe/H]. In our model, considering the same range of host star [Fe/H], the water mass fraction only decreases slightly from $\sim$15\% to $\sim$14\% when excluding iron oxidation and from $\sim$14\% to $\sim$10\% when including iron oxidation\footnote{These water mass fractions are calculated outside of the CO$_2$ ice line, where the water mass fraction is lowest. Inside the CO$_2$ ice line, the water mass fractions range from 21\% to 19\% when excluding iron oxidation and 20\% to 14\% when including iron oxidation.}.

\citet{adibekyan2021_comphost} used interior models of observed planets to determine their iron mass fraction (roughly equivalent to our core mass fractions) as a function of the iron-to-silicate ratio of their host stars. They found that the iron mass fraction of planets is expected to increase with the iron-to-silicate ratio of their host star. This result holds true no matter if they assume that all iron sits in the core or if some of the iron resides in the mantle of the planets. These results are therefore consistent with our model that excludes iron oxidation in the protoplanetary disc, an assumption which is in agreement with the slow kinetics of the oxidation reaction and diffusion \citep{grossman2012_ironoxidisation,johansen2022_mercury}.
\subsection{Forming giant planets around solar-like stars}\label{ssec:giants_around_suns}
Giant planets start forming at around [Fe/H] = 0 using our nominal values for fragmentation velocity $\vf$ (2 m/s) and dust diffusion coefficient $\delta_{\rm t}$ ($10^{-4}$), which agrees with observations as giant planets around solar metallicities and below are rare \citep{santos2001_giantocc,johnson2010_giantocc,thorngren2016_giantsfeh}. By decreasing $\delta_{\rm t}$ or increasing $\vf$, we can make pebble accretion much more efficient and therefore more easily form giant planets for lower [Fe/H] as well. This shows that it is possible to form giant planets around stars similar to the Sun in metallicity as well as slightly less metal-rich stars.

\begin{figure}
    \centering
    \includegraphics[width=\linewidth]{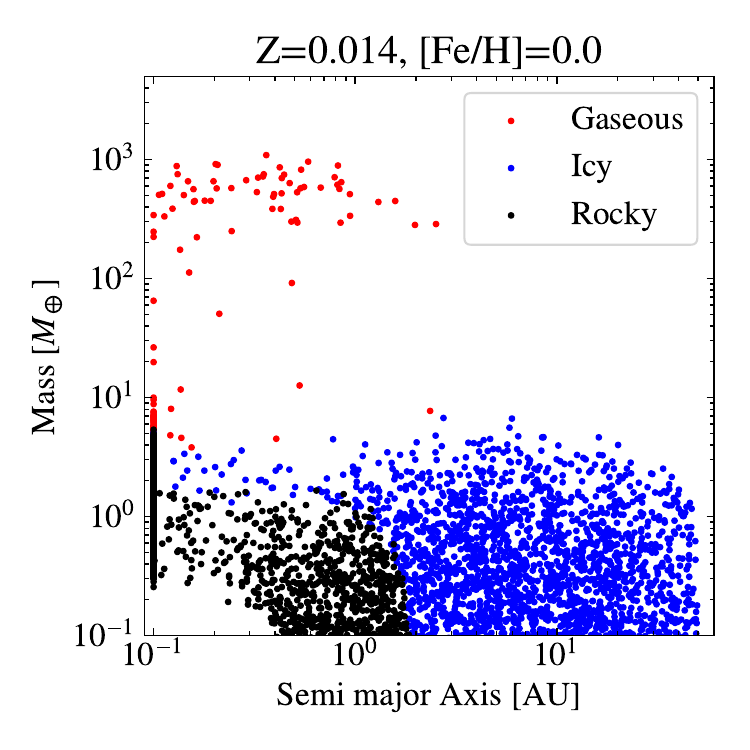}
    \caption{Resulting planet population around a star with solar composition and gas mass of 0.12 $M_\odot$. With this, more massive gas disc, we extend to lifetime of the disc as well as its size to 2.5 Myr and 73 AU respectively. This allows us to easily form giant planets further out in the disc compared to the middle panel of figure \ref{fig:MR_all}.}
    \label{fig:MR_largedisc}
\end{figure}

Further, given our disc model, we set a gas mass of 0.1 $M_\odot$ for a solar mass star and thus find a disc size of $\sim$60 AU. This is in line with observations, which have found dust and gas disc sizes to be typically below 100 AU \citep{ansdell2018_dustsizes,long2019_discsizes}. Higher gas masses would increase the disc lifetime and the disc size, thus allowing for the formation of giant planets further out in the disc. To show this, we simulate planet formation around a star with solar composition and a gas mass at 0.12 $M_\odot$ similar to the disc model in \citet{johansen2019}. This disc extends out to 73 AU and has a lifetime of $\sim$2.5 Myr. We show the resulting planet population in figure \ref{fig:MR_largedisc}. The result is similar to that of the middle panel in figure \ref{fig:MR_all} but giant planets now end in orbits out to $\sim$3 AU. Therefore it is possible that the Sun might have hosted a more massive protoplanetary disc than what we find using our disc model, which allowed it to form both Jupiter and Saturn.
\subsection{Systematic abundance uncertainty}
\label{ssec:abun_uncert}
In this study, we assume that the abundances of all $\alpha$ (O, Mg, Si, and S) scale with metallicity in the same way, following [Mg/Fe] for our stars. This is a reasonable assumption because all these elements are mainly produced in SNe II \citep[e.g.][]{timmes1995, rybizki2017} and similar distributions are seen in observational data \citep[e.g.][]{Bensby2014, Jonsson2020, Buder2021}. Carbon has a contribution from AGB stars \citep{busso1999_cprod} indicating that a different scaling relation might be required for C, however, in both GALAH and APOGEE Galactic surveys the behaviour of [C/Fe] closely matches the behaviour of [O/Fe], hence this assumption is sufficient for our purposes. \citet{delgadomena2021_coratio} suggest that for $\alpha$-rich and halo stars the C/O ratio is decreasing with decreasing [Fe/H]. However, for $\alpha$-poor stars, the C/O ratio is mostly flat\footnote{There is a large spread in [C/O] with a large uncertainty in the determination of the carbon abundance making it difficult to determine a clear relation between [C/O] and [Fe/H] for the $\alpha$-poor stars.}, agreeing with our scaling relation. Similar relations are also found by \citet{brewerfischer2016_COvFeh}. A lower C/O ratio for more [Fe/H]-poor stars, such as the stars in the halo and the $\alpha$-rich sample, leads to increased availability of oxygen to form water \citep{bitsch2020_model}. However, given that the pebble flux through protoplanetary discs around metal-poor stars is low, it is not clear that a different C/O ratio would have an effect on the number of water-rich planets formed around $\alpha$-rich stars and halo stars. The few water-rich planets formed, however, would be more water-rich if the C/O ratio of their host star would be lower.
\subsection{Different carbon model}\label{ssec:disc_chem_model}
In chemical equilibrium, CO is completely converted to CH$_4$ due to CO reacting with H$_2$ \citep{madhusudhan2012_carbon,madhusudhan2014_carbonmodel}. Carbon molecules, however, are not necessarily expected to be in equilibrium in protoplanetary discs. This makes it difficult to develop a concrete carbon model without including a fully time-dependent chemical network which would be computationally expensive. We, therefore, used a carbon model motivated by observations \citep{draine2003_carbongrains,pontoppidan2006_COobs,oberg2011_carbon}.

The hosts of carbon atoms not carried in CO and CO$_2$ were assumed to be refractory carbon grains due to the kinetic inhibition of CH$_4$ formation from CO in protoplanetary discs \citep{lodders2003_temps}. Another possible source of carbonaceous material could be in refractory organics which would sublimate at around 350 K \citep{lodders2004_organics}. Should a part of the carbon in volatile carriers instead be carried by carbon organics, the dust-to-gas ratio in the protoplanetary disc at regions below 350 K would increase, potentially allowing for more efficient planet formation. Further, the planets formed within this organics line would contain more carbon, lowering their core mass fractions.

Previous authors \citep{madhusudhan2014_carbonmodel,madhusudhan2017_model,bitsch2020_model,bitsch2018} have also included a second model for carbon-based on theoretical work by \citet{woitke2009_carbon}, which includes only CO, CO$_2$, and CH$_4$. It also introduces a break at the CO$_2$ ice line (70K) where outside of it, no CH$_4$ exists and all carbon sits in CO$_2$ and CO. Inside of the CO$_2$ ice line, half of the CO converts into CH$_4$ from the reaction 
\begin{equation}
    {\rm CO} + 3{\rm H}_2 = {\rm CH}_4 + \water.
\end{equation}
The abundance of the main oxygen carrier of the carbon molecules in our model, CO, is reduced in the second model (0.65C/H to 0.45C/H in the model with no grains) which slightly boosts the mass fraction of water in the disc. It is clear that a better understanding of carbon chemistry in protoplanetary discs and to what degree carbon chemistry is in equilibrium in protoplanetary discs would be needed for a fully consistent model of the accretion of carbon-bearing species during planet formation.
\subsection{Including multiple planets}\label{ssec:multiple_planets}
We do not take into account the possible gravitational interactions between protoplanets and grown planets in our model setup. Including the effects of multiple planets could have a major effect on the final planet population, as the formation of a massive planet could reduce the pebble flux interior of its orbit \citep{lambrechts2014_gasacc,mulders2021_MdwarfSE,liu2022_separation} although the efficiency of which a giant planet blocks the pebble flux has been shown to be limited when taking into account the fragmentation of pebbles \citep{stammler2023_fluxblock}. Further, even with a low pebble flux, it would still be possible for protoplanets to grow into Earth-mass planets through mutual collisions, similar to classical terrestrial planet formation \citep{morbidelli2012_formreview}. For higher pebble fluxes, growth and subsequent migration are fast, causing protoplanets to pile up in the inner regions of the disc where mutual collisions can drive the formation of super-Earths \citep{lambrechts2019_multiple}. The compositions of planets would also be expected to be different when including gravitational effects between protoplanets as collisions could increase the heavy metal contents of giant planets \citep{ogihara2021_composition_multiple}.
\section{Summary and conclusions}
\label{sec:conclusion}
In this work, we combine the observations of Galactic stars with the planet growth and disc models in order to understand how the formation of planets depends on the parameters of stars in the context of Galactic structure, specifically stellar mass, metallicity [Fe/H], and $\alpha$-enhancement. We employ the NLTE chemical abundances and masses of stars from the analysis of the Gaia-ESO survey data by \citet{Gent2022b}. By using a simple 1-D $\alpha$-disc model for the protoplanetary discs and applying a chemical model including refractory minerals (based on equilibrium chemistry) and volatile ices (based on observations), we have been able to track final masses, orbits and compositions of the planets formed. We could therefore quantify the characteristic differences in the properties (mass, type) of planets forming in the halo, thin, and thick disc of the Milky Way galaxy.

From this work, we have been able to draw the following conclusions:

\begin{enumerate}

    \item We have found that stars with increasing [Fe/H] form planets with increasing masses for both giant planets and rocky planets. This trend is also seen observationally \citep{buchhave2012_plmassvsfeh,swastik2022_hostchem}.

    \item Both stellar mass and metallicity [Fe/H] affect the formation of giant planets and super-Earths. Higher [Fe/H] results in more water-rich planets due to the increase in mass accretion rate of protoplanets growing beyond the water ice line. Higher stellar mass results in the formation of more massive protoplanetary discs, capable of sustaining a high pebble flux for a long time. This causes water-rich planets to grow quickly enough to accrete a massive gaseous envelope, resulting in the fraction of water-rich planets decreasing slightly with higher stellar mass. 
 
    \item Super-Earths, giant planets, gaseous planets, and water-rich planets are all more common around stars in the $\alpha$-poor (thin disc) sample compared to the $\alpha$-rich (thick disc) sample due to the overall higher metallicities of the $\alpha$-poor stars while rocky, Earth-mass planets dominate the planet population around halo stars. The fraction of rocky and Earth-mass planets decreases with increasing [Fe/H] as the planet population becomes more diverse. We caution that in this work the distinction between the thin and thick discs is based on stellar [$\alpha$/Fe] ratios, following a common approach in the literature, however in future, more detailed, and complete studies of the Galactic disc, such as the 4MIDABLE surveys on 4MOST \citep{Chiappini2019, Bensby2019} will yield a more precise definition of the properties of this Galactic component.

    \item We compare our results using a chemical model including iron oxidation with one excluding iron oxidation. Including iron oxidation, the core mass fractions of planets around $\alpha$-poor stars and $\alpha$-rich stars decrease with increasing [Fe/H]. When excluding iron oxidation, the core mass fractions instead increase with increasing [Fe/H], since the amount of available iron (which increases with [Fe/H]) is in this case partitioned into the core instead of in the mantle in the form of the lithophile minerals fayalite and magnetite. Further, when excluding iron oxidation, we are able to match the core mass fraction of the Earth while we do not form any planets with a low enough core mass fraction to match the core mass fraction of Mars. We attribute this to our simple estimation of core mass fraction not taking into account the possible oxidation of iron in contact with accreted water ice \citep{johansen2023_cmftheory}. Including iron oxidation results in low core mass fractions, almost matching that of Mars. However, we are only able to form a handful of planets with a similar core mass fraction to that of Earth. These results imply that iron oxidation does not occur in protoplanetary discs, in agreement with the slow kinetics of the oxidation reaction and diffusion \citep{grossman2012_ironoxidisation,johansen2022_mercury}.
\end{enumerate}

Overall, we believe that our study contributes towards understanding the role of Galactic chemical evolution in shaping the planet populations of the Milky Way and that it will be useful in order to be able to interpret and predict results from upcoming exoplanet surveys such as PLATO \citep{rauer2014_plato}.

\begin{acknowledgements}
We would like to thank the anonymous referee for their insightful comments and suggestion which helped improve the manuscript. A.J. acknowledges funding from the European Research Foundation
(ERC Consolidator Grant 724687-PLANETESYS), the Knut and Alice Wallenberg Foundation (Wallenberg Scholar Grant 2019.0442), the Swedish Research Council (Project Grant 2018-04867), the Danish National Research Foundation (DNRF Chair Grant DNRF159) and the Göran Gustafsson Foundation. 
MB and MG are supported through the Lise Meitner grant from the Max Planck Society. We acknowledge support by the Collaborative Research centre SFB 881 (projects A5, A10), Heidelberg University, of the Deutsche Forschungsgemeinschaft (DFG, German Research Foundation). This project has received funding from the European Research Council (ERC) under the European Union’s Horizon 2020 research and innovation programme (Grant agreement No. 949173).

\end{acknowledgements}
\bibliographystyle{aa}
\bibliography{bib}
\begin{appendix}
\section{Calculating the temperature at the RCB}
\label{asec:T_rcb}
The pressure at the $R_{\rm RCB}$ is given by 
\begin{equation}\label{eq:p_rcb}
    P_{\rm RCB} = P_0\frac{\nabla_{\rm ad}/\nabla_0-\nabla_{\rm ad}/\nabla_{\infty}}{1-\nabla_{\rm ad}/\nabla_\infty},
\end{equation}
where $P_0$ is the pressure at the outer boundary (i.e. of the gas in the disc). The remaining quantities are defined as follows
\begin{equation}\label{eq:nabla}
    \begin{split}
        & \nabla_{\rm ad} = \frac{\gamma_{\rm ad}-1}{\gamma_{\rm ad}} \\
        & \nabla_\infty = \frac{1}{4-\beta_{\rm op}} \\
        & \nabla_0 = \frac{3\kappa P_0}{64\pi G M \sigma_{\rm SB} T_0^4}L,
    \end{split}
\end{equation}
where $\gamma_{\rm ad} = 1.4$ is the adiabatic index of the envelope gas, $\beta_{\rm op}=2$ is the opacity index for the disc, $M$ is the mass of the planet, $\sigma_{\rm SB}$ is the Stefan-Boltzmann constant, $T_0$ is the disc temperature at the outer boundary, and $L$ is the luminosity of the planet defined as
\begin{equation}\label{eq:lum_pl}
    L = \frac{GM\Dot{M}}{R}.
\end{equation}
In order to calculate the sizes $R$ of the planets, we assume that they are spherical bodies, meaning that
\begin{equation}\label{eq:R_pl}
    R = \left(\frac{3M}{4\pi\rho}\right)^{1/3},
\end{equation}
where $\rho$ is the uncompressed bulk density of the solid component of the planets. The temperature at the RCB can then be written as
\begin{equation}\label{eq:T_rcb}
    T_{\rm RCB} = T_0\left(\frac{\nabla_0}{\nabla_\infty}\left[\frac{P_{\rm RCB}}{P_0}-1\right]+1\right)^{\nabla_\infty}.
\end{equation}
\section{Mass fraction of solids for different values of [Fe/H]}
\label{asec:solid_fraction_feh}
\begin{figure*}
    \includegraphics[width=\linewidth]{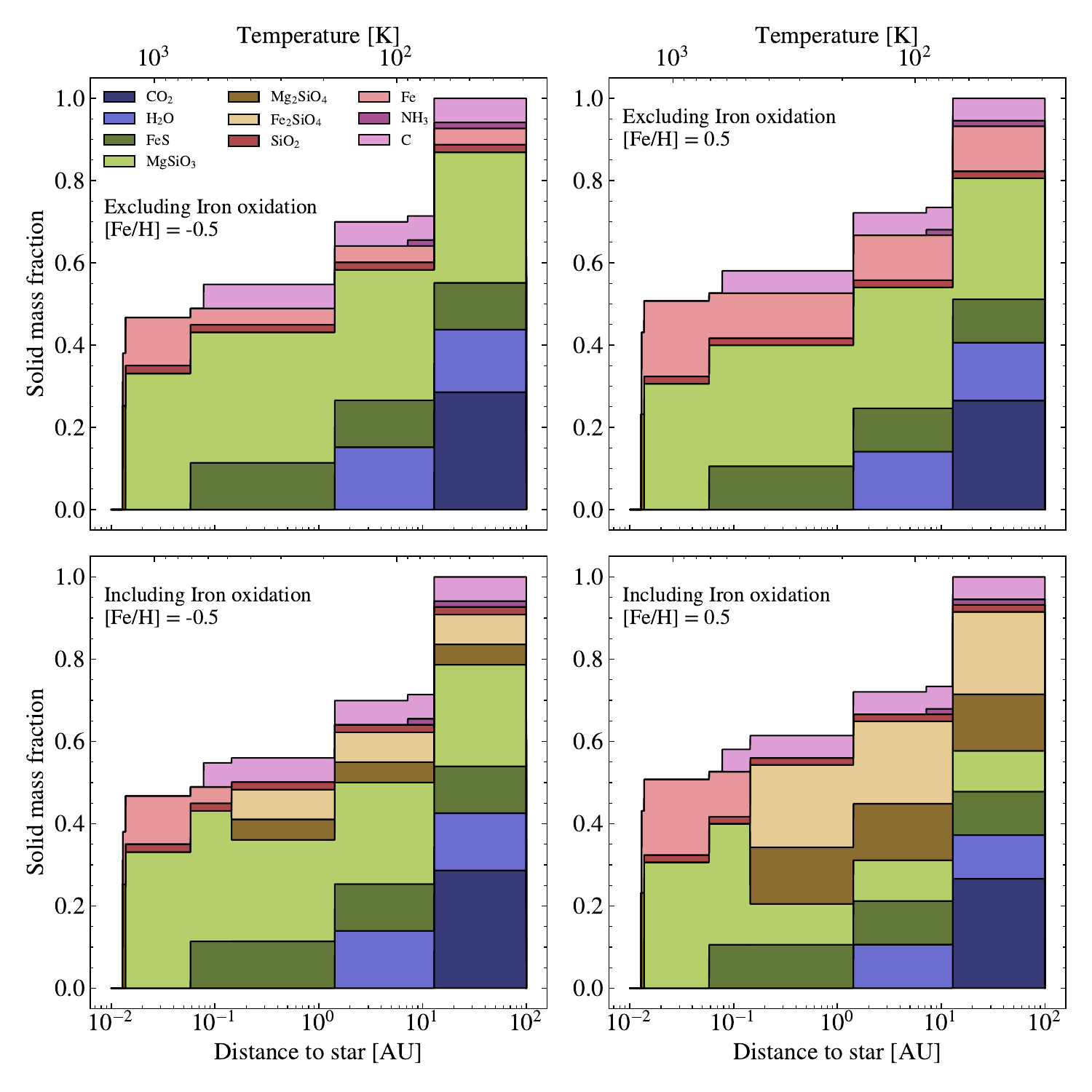}
    \caption{Same as figure \ref{fig:solid_fraction} but for $\feh = -0.5$ (left) and $\feh = 0.5$ (right). The water content in the disc changes very little with [Fe/H] when excluding iron oxidation as discussed in section \ref{ssec:compare} while a slight increase in the water mass fraction can be seen for the metal-poor star when including iron oxidation. Further, due to the increased iron content, the mass fraction of fayalite (and forsterite from the reaction shown in \eqref{eq:fayalite}) is increased in the metal-rich star.}
    \label{fig:solid_fraction_differentfeh}
\end{figure*}

In figure \ref{fig:solid_fraction_differentfeh} we show the mass fractions of solids in the disc for a synthetic metal-poor star (left) and a synthetic metal-rich (right) star where the abundances has been scaled according to the fit in figure \ref{fig:mgfe_mgh_feh}. When excluding iron oxidation, the mass fraction of metallic Fe increases from a $\sim$5\% to $\sim$10\% while the mass fractions of the remaining species remain relatively unchanged as the relative abundance values of all elements other than Fe remain constant. When including iron oxidation, the same effect is seen for fayalite. The metal-rich star has a slight decrease in the water mass fraction as well as the enstatite mass fraction while the forsterite mass fraction is increased due to the reaction shown in \eqref{eq:fayalite}. Both when including and excluding iron oxidation, the mass fraction of FeS is relatively unchanged due it being limited by the ratio S/Fe which remains similar for these values of [Fe/H].

\end{appendix}
\end{document}